\documentclass{article}
\usepackage{latexsym}
\usepackage{amsmath}
\usepackage{amssymb}
\usepackage{amsfonts}
\usepackage{amsxtra}
\usepackage{cite}
\allowdisplaybreaks
\renewcommand{\P}{\mathcal{P}}
\newcommand{\ver}[1]{\widehat{#1}}
\newcommand{\verp}[1]{\ver{p}^{\,#1}}
\newcommand{\verr}[1]{\ver{r}^{\,#1}}
\newcommand{\verrp}[1]{\widehat{r}\,'^{#1}}
\newcommand{\verpp}[1]{\widehat{p}\,'^{#1}}
\newcommand{\rdotrp}{\verr{}\cdot\verrp{}}
\newcommand{\pdotpp}{\verp{}\cdot\verpp{}}
\newcommand{\ppdotq}{\verpp{}\cdot\verq{}}
\newcommand{\pdotq}{\verp{}\cdot\verq{}}
\newcommand{\verq}[1]{\ver{q}^{\,#1}}

\newcommand{\vchi}[1]{\ver{\chi}_{#1}}
\newcommand{\vchiU}[2]{\ver{\chi}_{#1}^{\,#2}}
\newcommand{\veps}[1]{\ver{\varepsilon}^{\,#1}}
\newcommand{\CG}[6]{\langle #1,#2;#3,#4 | #5,#6\rangle}
\newcommand{\SJ}[6]{\left\{\begin{array}{ccc}
#1 & #2 & #3\\ #4 & #5 & #6 \end{array}\right\}}

\newcommand{\lave}{\langle\ell\rangle}
\newcommand{\dl}{\Delta\ell}
\newcommand{\save}{\langle s\rangle}
\newcommand{\ds}{\Delta s}

\title{Addition theorems for spin spherical harmonics. I
  Preliminaries} 
\author{Antonio O.\ Bouzas \thanks{E-mail:
    abouzas@mda.cinvestav.mx}\\\small Departamento de F\'{\i}sica
  Aplicada, CINVESTAV-IPN \\\small Carretera Antigua a Progreso Km.\
  6, Apdo.\ Postal 73 ``Cordemex''\\\small
  M\'erida 97310, Yucat\'an, M\'exico}
\begin{document}
\maketitle
\begin{abstract}
We develop a systematic approach to deriving addition theorems for,
and some other bilocal sums of, spin spherical harmonics.  In this
first part we establish some necessary technical results.  We discuss
the factorization of orbital and spin degrees of freedom in certain
products of Clebsch-Gordan coefficients, and obtain general explicit
results for the matrix elements in configuration space of tensor
products of arbitrary rank of the position and angular-momentum
operators.  These results are the basis of the addition theorems for
spin spherical harmonics obtained in part II.
\end{abstract}

\section{Introduction}
\label{sec:intro}

The importance of the representation theory of the three-dimensional
rotation group \cite{edm96,wig59,jud75,bie85a,bie85b,var88,ham89} in
the study of all natural quantum systems hardly needs to be mentioned.
That central role is due to the fact that the four fundamental
interactions of nature, and their effective interactions relevant to
nuclear, atomic and molecular physics, are all invariant under the
rotation group.  It is also for that reason that partial-wave
expansions are essential tools in the analysis of both classical and
quantum scattering processes.

The algebraic aspects of the computation of partial-wave expansions
with spin states are best codified by the addition theorems for spin
spherical harmonics.  Spin-$s$ spherical harmonics $Y^{\ell
  s}_{jj_z}(\verr{})$ \cite{bie85a,var88} are the angular-momentum
eigenfunctions relevant to the description of spin-$s$ particles
subject to spin-dependent central interactions.\footnote{Familiar
  examples of spin-$s$ spherical harmonics are ordinary scalar
  spherical harmonics, $Y^{\ell 0}_{\ell\ell_z}(\verr{}) =
  Y_{\ell\ell_z}(\verr{})$, and vector spherical harmonics
  \cite{bie85a,var88,gal90,jackxx}, $Y^{\ell 1}_{j j_z}(\verr{})$.}
As such, they are of interest in their own right beyond their
above-mentioned specific application to partial-wave expansions that
constitutes our main motivation for their study.  In this paper,
together with its second part \cite{bou10} (hereafter referred to as
II), we develop a systematic framework to derive addition theorems for
spin spherical harmonics, and employ it to obtain those addition
theorems for low spins, $0<s', s\leq 3/2$, and for arbitrary integer
spin $s'$ if $s=0$.
 
In this first part we establish the preliminary results that will be
used as basic building blocks for constructing addition theorems for
spin spherical harmonics in II.  We consider first the factorization
of orbital and spin degrees of freedom in products of Clebsch-Gordan
(henceforth CG) coefficients of the form
$\CG{\ell'}{\ell'_z}{s'}{s'_z}{j}{j_z}
\CG{\ell}{\ell_z}{s}{s_z}{j}{j_z}$.  We obtain a general factorization
result, make its tensor and spinor structure explicit, and discuss the
particular cases relevant to spins 1/2, 1 and 3/2.  Such products of
CG coefficients occur in the computation of physical quantities, such
as scattering amplitudes or other matrix elements, related to
transitions among states with definite orbital and spin quantum
numbers.  Thus, these results may be of interest independently of the
addition theorems considered here and in II. We study also the matrix
elements of arbitrary tensor products of orbital operators between
angular-momentum projections of position eigenstates, for which we
obtain completely general results.  These are the other main
ingredient in our derivation of addition theorems for spin spherical
harmonics in II.  Those matrix elements are closely related to bilocal
spherical harmonics and other bilocal sums of ordinary spherical
harmonics, as discussed in detail in II, so we expect that these
results and the approach used to derive them should be applicable in
other contexts as well.

The outline of the paper is as follows.  In the next section we
discuss the factorization of orbital and spin degrees of freedom in
products of CG coefficients.  In section \ref{sec:proj} we
introduce the angular-momentum projector operator and obtain general
expressions for its matrix elements with orbital tensor operators.  In
appendix \ref{sec:notatio} we state our notation and conventions.  In
appendix \ref{sec:standard} we give the definitions and main
properties of the standard bases of irreducible tensors and spinors,
of arbitrary spin, used as spin wavefunctions throughout the paper.  A
detailed list of general results for reduced matrix elements of
arbitrary tensor products of orbital and spin operators, needed in the
main sections of the paper, are given in appendix \ref{sec:matele}.
Finally, appendix \ref{sec:appa} gathers some ancillary calculations
needed in sect.\ \ref{sec:proj}.

\section{Factorization of orbital and spin dependence}
\label{sec:fac}

We begin by considering products of CG coefficients of the form
\begin{equation}
  \label{eq:additionintro}
  \begin{aligned}
  S(j,\ell',\ell,s',s;\ell'_z,\ell_z,s'_z,s_z) &= \sum_{j_z=-j}^j 
  \CG{\ell'}{\ell'_z}{s'}{s'_z}{j}{j_z} 
  \CG{\ell}{\ell_z}{s}{s_z}{j}{j_z} \\
  &=
  \CG{\ell'}{\ell'_z}{s'}{s'_z}{j}{\ell'_z+s'_z} 
  \CG{\ell}{\ell_z}{s}{s_z}{j}{\ell_z+s_z}
  \delta_{\ell'_z+s'_z,\ell_z+s_z}~,
  \end{aligned}
\end{equation}
with $\ell'$, $\ell$ integer and $s'$, $s$ integer or half-integer.
We will usually write $S(j,\ell',\ell,s',s)$, omitting the last
arguments for brevity.  For the purpose of obtaining addition theorems
for spin spherical harmonics we need to rewrite $S(j,\ell',\ell,s',s)$
as a sum of terms with a completely factorized dependence on orbital
and spin angular-momentum projections, and to make  explicit the
tensor structure of those terms.  The first goal is achieved by making
use of relation (\ref{eq:6j-1}) to write (see appendix
\ref{sec:notatio} for our notation and conventions),
\begin{equation}
  \label{eq:6j-2}
  \begin{gathered}
  \begin{aligned}
  S(j,\ell',\ell,s',s) &= 
  (-1)^{s'_z+s_z}(-1)^{\ell+s'-j}
  \frac{2j+1}{\sqrt{(2\ell'+1)(2s'+1)}}
    \sum_{\Delta=\Delta_\mathrm{min}}^{\Delta_\mathrm{max}} (2\Delta+1)
  \SJ{\ell'}{\Delta}{\ell}{s}{j}{s'}
  \\
  &\quad\times
  \CG{\ell}{\ell_z}{\Delta}{\Delta\ell_z}{\ell'}{\ell'_z}
  \CG{s}{s_z}{\Delta}{\Delta s_z}{s'}{s'_z}~,
  \end{aligned}\\
  \Delta_\mathrm{min} = \max\{|\Delta\ell|,|\Delta s|\}~,
  \qquad
  \Delta_\mathrm{max} = \min\{\ell'+\ell,s'+s\}~,
  \qquad  
  \dl_z + \ds_z =0~.
  \end{gathered}
\end{equation}
In order to make the tensor structure of each term on the r.h.s.\
explicit, we use the Wigner-Eckart (henceforth WE) theorem 
\cite{edm96,wig59,jud75,bie85a,bie85b,var88,gal90} to write
the CG coefficients in terms of appropriate irreducible tensor
operators,
\begin{equation}
  \label{eq:CGmaster}
  \begin{gathered}
  \begin{aligned}
  S(j,\ell',\ell,s',s) &=
  \sum_{\Delta=\Delta_\mathrm{min}}^{\Delta_\mathrm{max}}
  C^{s's\Delta}_{\ell'\ell j} \langle\ell',\ell'_z | \veps{i_1\ldots
    i_\Delta}(\Delta\ell_z) \verr{i_1}\ldots \verr{i_{|\Delta\ell|}}
  L^{i_{|\Delta\ell|+1}} \ldots L^{i_\Delta} | \ell, \ell_z\rangle\\
  &\quad\times
  \langle s',s'_z | \veps{h_1\ldots
    h_\Delta}(\Delta\ell_z)^* T^{h_1}\ldots T^{h_{|\Delta s|}}
  S^{h_{|\Delta s|+1}} \ldots S^{h_\Delta} | s, s_z\rangle  ~,
\end{aligned}\\
\begin{aligned}
C^{s's\Delta}_{\ell'\ell j} &= (-1)^{2s'_z}(-1)^{\ell+s'-j}
\frac{2j+1}{\sqrt{(2\ell'+1) (2s'+1)}} (2\Delta+1)
\SJ{\ell'}{\Delta}{\ell}{s}{j}{s'} \\
&\quad\times \left(\langle\ell'|| \veps{j_1\ldots
    j_\Delta} \verr{j_1}\ldots \verr{j_{|\Delta\ell|}}
  L^{j_{|\Delta\ell|+1}} \ldots L^{j_\Delta} || \ell\rangle
 \langle s'|| \veps{k_1\ldots
    k_\Delta} T^{k_1}\ldots T^{k_{|\Delta s|}}
  S^{k_{|\Delta s|+1}} \ldots S^{k_\Delta} || s\rangle
  \right)^{-1}~,
\end{aligned}
  \end{gathered}
\end{equation}
where $\Delta_{\mathrm{min},\mathrm{max}}$ are as in (\ref{eq:6j-2}),
and our notation for the standard basis tensors $\veps{i_1\ldots i_n}$
and for irreducible tensor operators is explained in appendix
\ref{sec:standard}.  The operators appearing in (\ref{eq:CGmaster})
are the orbital $\vec{L}$ and spin $\vec{S}$ angular momenta, the
position versor $\verr{}=\vec{r}/r$, and the spin-transition operator
$\vec{T}$ defined here as the spin analog of $\verr{}$ (see sect.\
\ref{sec:spintrans} below for a detailed discussion of this operator).
The reduced matrix elements appearing in (\ref{eq:CGmaster}) are
explicitly given for arbitrary values of their parameters in appendix
\ref{sec:matele}.

Whereas (\ref{eq:CGmaster}) is completely general, it is not yet
explicit enough for the purpose of deriving addition theorems for spin
spherical harmonics.  From a practical point of view, we need only
consider low values of $s'$, $s$ since quantum states with high values
of spin occur infrequently in nature.  In this paper we restrict
ourselves to $s'$, $s=0$, 1/2, 1, 3/2, and $|\Delta s|\leq 1$.  For
those spins the coefficients $C^{s's\Delta}_{\ell'\ell j}$ can be
drastically simplified by exploiting the fact that $j-\ell'$ and
$j-\ell$ can take only a small set of values.  The spin matrix
elements in (\ref{eq:CGmaster}) can be compactly evaluated in terms of
standard tensors and spinors, and spin matrices. (The orbital matrix
elements could, in principle, also be expressed in that way, but it
would be impractical because $\ell'$, $\ell$ are not bounded above
thus requiring tensor wave functions of arbitrarily large rank.  We
evaluate those matrix elements with a completely different technique
below in section \ref{sec:proj}.)  Clearly, we must discuss separately
the different values of $s'$ and $s$, as those will determine the kind
of spin spherical harmonics involved in the addition theorem.  For
fixed $s'$ and $s$, different values of $\Delta\ell$ correspond to
different addition theorems, so we must consider those cases
separately as well.  Furthermore, both in addition theorems for spin
spherical harmonics and in partial wave expansions, terms with
different $|\Delta\ell|$ have different tensor and spin structure.

\subsection{$\boldsymbol{s'=1/2=s}$}
\label{sec:1half}

For $s'=1/2=s$ there are two possible values of $|\Delta\ell|=0$, 1.

\paragraph*{$\boldsymbol{\Delta\ell =0}$}
\label{sec:1half-0}

In the case $s=1/2$, $j$ can take only two values and
(\ref{eq:CGmaster}) reduces to 
\begin{equation}
  \label{eq:1half-1}
  \begin{gathered}
S(j,\ell,\ell,1/2,1/2) = C^{\frac{1}{2}\frac{1}{2}0}_{\ell\ell j} \delta_{\ell'_z \ell_z}
\delta_{s'_z s_z} + C^{\frac{1}{2}\frac{1}{2}1}_{\ell\ell j} \langle \ell,\ell'_z |
\veps{k}(\dl_z) L^k | \ell,\ell_z \rangle\, \veps{h}(\dl_z)^*
\vchi{A}(s'_z)^* \frac{1}{2}\sigma^h_{AB}\vchi{B}(s_z) ~,\\
  C^{\frac{1}{2}\frac{1}{2}0}_{\ell\ell j} = \frac{j+1/2}{2\ell+1}~,
\qquad
  C^{\frac{1}{2}\frac{1}{2}1}_{\ell\ell j} =
  \frac{2(j-\ell)}{\ell+1/2}~,
  \end{gathered}
\end{equation}
which is a well-known result (see e.g., \cite{nac90}).  Notice that,
due to the conservation of $j_z$, the product
$\veps{k}(\dl_z)\veps{h}(\dl_z)^*$ in (\ref{eq:1half-1}) can be replaced
by $\delta^{kh}$.  We quote here also the simplified form of
(\ref{eq:6j-2}) in this case,
\begin{equation}
  \label{eq:1half-2}
  S(j,\ell,\ell,1/2,1/2) = C^{\frac{1}{2}\frac{1}{2}0}_{\ell\ell j} \delta_{\ell'_z \ell_z}
  \delta_{s'_z s_z} + \frac{\sqrt{3}}{2} \sqrt{\ell(\ell+1)} C^{\frac{1}{2}\frac{1}{2}1}_{\ell\ell j}
  \CG{\ell}{\ell_z}{1}{\dl_z}{\ell}{\ell'_z}
  \CG{1/2}{s'_z}{1}{\dl_z}{1/2}{s_z}~,
\end{equation}
which results from (\ref{eq:1half-1}) by applying the WE theorem.
 
\paragraph*{$\boldsymbol{|\Delta\ell|=1}$}
\label{sec:1half-1}

In this case $S(j,\ell',\ell,1/2,1/2)$ can be non-vanishing only if
$\ell'=\ell+1$ and $j=\ell+1/2$, or if $\ell'=\ell-1$ and
$j=\ell-1/2$.  Equation (\ref{eq:CGmaster}) takes the simplified form,
\begin{equation}
  \label{eq:1half-3}
S(j,\ell_1,\ell,1/2,1/2) = C^{\frac{1}{2}\frac{1}{2}1}_{\ell_1\ell j}
\langle \ell_1,\ell'_z | \veps{k}(\dl) \verr{k} | \ell,\ell_z \rangle\,
\veps{h}(\dl)^* \vchi{A}(s'_z)^* 
  \frac{1}{2}\sigma^h_{AB}\vchi{B}(s_z) ~,
\quad
  C^{\frac{1}{2}\frac{1}{2}1}_{\ell_1\ell j} = -2~,
\end{equation}
which is appropriate to derive an addition theorem for spin-1/2
spherical harmonics.  As in the previous case, the product of standard
versors can be substituted by $\delta^{kh}$. Similarly,
(\ref{eq:6j-2}) reduces to
\begin{equation}
  \label{eq:1half-4}
  S(j,\ell_1,\ell,1/2,1/2) = -\dl\sqrt{\frac{3}{2}}
  \sqrt{\frac{2\lave+1}{2\ell_1+1}}
  \CG{\ell}{\ell_z}{1}{\Delta\ell_z}{\ell_1}{\ell'_z} 
  \CG{1/2}{s'_z}{1}{\Delta\ell_z}{1/2}{s_z} ~.
\end{equation}
As a side remark, we notice that in this case  $(2\lave+1)/(2\ell_1+1)
= (2j+1)/(2j+1+\dl)$, which provides an alternate form for
(\ref{eq:1half-4}). 

\subsection{$\boldsymbol{s'=1=s}$}
\label{sec:1}

The general result (\ref{eq:CGmaster}) is greatly simplified in the
case of $S(j,\ell',\ell,1,1)$ because, for fixed $\ell$, $j-\ell$ can
take only the three values $\pm1$, 0. The spin matrix elements are
given by
\begin{equation}
  \label{eq:1-1}
  \begin{aligned}
  \langle 1,s'_z | \veps{h}(\dl_z)^* S^h | 1,s_z\rangle = -i
  \veps{h}(\dl_z)^* \left(\veps{}(s'_z)^* \wedge
    \veps{}(s_z)\right)^h~,  \\
  \langle 1,s'_z | \veps{h_1h_1}(\dl_z)^* S^{h_1}S^{h_2} |
  1,s_z\rangle = -\veps{h_1h_2}(\dl_z)^* \veps{h_1}(s'_z)^*
  \veps{h_2}(s_z)~.  
  \end{aligned}
\end{equation}
There are three possible values for $\Delta\ell=0$, 1, 2.

\paragraph*{$\boldsymbol{|\Delta\ell|=0}$}
\label{sec:1-1}

In this case (\ref{eq:CGmaster}) can be written as
\begin{equation}
  \label{eq:1-2}
  \begin{gathered}
  \begin{aligned}
S(j,\ell,\ell,1,1) &= C^{110}_{\ell\ell j} \delta_{\ell'_z\ell_z}
\delta_{s'_zs_z} - C^{111}_{\ell\ell j} \langle \ell,\ell'_z |
\veps{k}(\dl_z) L^k | \ell,\ell_z\rangle\,\veps{h}(\dl_z)^*
i \left(\veps{}(s'_z)^* \wedge \veps{}(s_z)\right)^h\\
&\quad - C^{112}_{\ell\ell j} \langle \ell,\ell'_z |
\veps{k_1k_2}(\dl_z) L^{k_1} L^{k_2} | \ell,\ell_z\rangle\,\veps{h_1h_2}(\dl_z)^*
\veps{h_1}(s'_z)^* \veps{h_2}(s_z)~,
  \end{aligned}\\
  C^{110}_{\ell\ell j} = \frac{1}{3} \frac{2j+1}{2\ell+1}~,
  \quad
  C^{111}_{\ell\ell j} = \frac{2}{D_{\ell j}} (2j+1)
  ((j-\ell)(j+\ell+1)+1)~,
  \quad
  C^{112}_{\ell\ell j} = \frac{4}{D_{\ell j}} (2j+1)~,\\
  D_{\ell j} = (-1)^{j-\ell-1} \left((j-\ell)^2+1\right)(j+\ell+2)
  (j+\ell+1) (j+\ell)~.
  \end{gathered}
\end{equation}
This form of (\ref{eq:CGmaster}) is needed in the derivation of
addition theorems for vector spherical harmonics.  Notice that in the
term multiplied by $C^{111}_{\ell\ell j}$ we can replace
$\veps{k}(\dl_z) \veps{h}(\dl_z)^*$ by $\delta^{kh}$, and in the term
multiplied by $C^{112}_{\ell\ell j}$ we can rewrite the matrix
elements as
\begin{equation}
  \label{eq:1-3}
  \langle \ell,\ell'_z | \veps{k_1k_2}(\dl_z)L^{k_1}L^{k_2} |
  \ell,\ell_z \rangle\, \veps{h_1h_2}(\dl_z)^*
  \veps{h_1}(s'_z)^* \veps{h_2}(s_z) =  
  \langle \ell,\ell'_z |\frac{1}{2}L^{\{p} L^{q\}_0} |\ell,\ell_z \rangle\,
  \veps{p}(s'_z)^* \veps{q}(s_z)~.
\end{equation}
The relation (\ref{eq:6j-2}) in this case reduces to
\begin{equation}
  \label{eq:1-4}
  \begin{aligned}
S(j,\ell,1) &=
C^{110}_{\ell\ell j}\delta_{\ell'_z\ell_z}\delta_{s'_zs_z}  
+\sqrt{2} \sqrt{\ell(\ell+1)}C^{111}_{\ell\ell j}
\CG{\ell}{\ell_z}{1}{\dl_z}{\ell}{\ell'_z}  
\CG{1}{s'_z}{1}{\dl_z}{1}{s_z} \\
&\quad + \frac{\sqrt{10}}{6} \sqrt{\ell(\ell+1)}
\sqrt{(2\ell-1)(2\ell+3)} C^{112}_{\ell\ell j}
\CG{\ell}{\ell_z}{2}{\dl_z}{\ell}{\ell'_z}  
\CG{1}{s'_z}{2}{\dl_z}{1}{s_z}~, 
  \end{aligned}
\end{equation}
with $C^{11\Delta}_{\ell\ell j}$ as in (\ref{eq:1-2}).

\paragraph*{$\boldsymbol{|\Delta\ell|=1}$}
\label{sec:1-2}

In this case $S(j,\ell',\ell,1,1)$ can be non-vanishing only if
$\ell'=\ell+1$ and $j=\ell+1$, $\ell$, or if $\ell'=\ell-1$ and
$j=\ell$, $\ell-1$.  With that discrete set of values for $j-\ell$,
the spin matrix elements (\ref{eq:1-1}), and the reduced matrix
elements in appendix \ref{sec:matele}, from (\ref{eq:CGmaster}) we
obtain
\begin{equation}
  \label{eq:1-5}
  \begin{gathered}
  \begin{aligned}
    S(j,\ell_1,\ell,1,1) &= -C^{111}_{\ell_1\ell j}
    \langle\ell_1,\ell'_z|\veps{k}(\Delta\ell_z)\verr{k}|\ell,\ell_z\rangle\, 
    i \veps{h}(\Delta\ell_z)^*
    (\veps{}(s'_z)^*\wedge\veps{}(s_z))^h\\
  &\quad - C^{112}_{\ell_1\ell j} 
    \langle\ell_1,\ell'_z|\veps{k_1k_2}(\Delta\ell_z)\verr{k_1}L^{k_2}|\ell,\ell_z\rangle
    \,
    \veps{h_1h_2}(\Delta\ell_z)^* \veps{h_1}(s'_z)^*\veps{h_2}(s_z)~,
  \end{aligned}\\
  C^{111}_{\ell_1\ell j} = -\frac{1}{2\lave+1} \sqrt{2j+1}
  \sqrt{2j-\lave+1/2}~,
  \quad
  C^{112}_{\ell_1\ell j} = -4 \frac{(j-\lave)}{2\lave+1}
  \sqrt{\frac{2j+1}{2j-\lave+1/2}} ~. 
  \end{gathered}
\end{equation}
As before, on the first line of (\ref{eq:1-5}) we can replace
$\veps{k}(\dl_z) \veps{h}(\dl_z)^*$ by $\delta^{kh}$, and on the
second line we can rewrite the matrix elements as
\begin{equation}
  \label{eq:1-6}
  \langle \ell',\ell'_z | \veps{k_1k_2}(\dl_z)\verr{k_1}L^{k_2} |
  \ell,\ell_z \rangle\, \veps{h_1h_2}(\dl_z)^*
  \veps{h_1}(s'_z)^* \veps{h_2}(s_z) =  
  \langle \ell',\ell'_z |\frac{1}{2}\verr{\{p} L^{q\}_0} |\ell,\ell_z \rangle\,
  \veps{p}(s'_z)^* \veps{q}(s_z)~.
\end{equation}
The form of (\ref{eq:6j-2}) in this case is
\begin{equation}
  \label{eq:1-7}
  \begin{aligned}
    S(j,\ell_1,\ell,1,1) &=
    \dl \sqrt{\frac{2\lave+1}{2\ell_1+1}} \left(
      \rule{0pt}{18pt}
      C^{111}_{\ell_1\ell j}
    \CG{\ell}{\ell_z}{1}{\Delta\ell_z}{\ell_1}{\ell'_z} 
    \CG{1}{s'_z}{1}{\Delta\ell_z}{1}{s_z}  
    \right.\\
    &\quad \left. +\frac{1}{4}\sqrt{\frac{5}{3}} 
    \sqrt{(2\lave-1)(2\lave+3)} C^{112}_{\ell_1\ell j}
    \CG{\ell}{\ell_z}{2}{\Delta\ell_z}{\ell_1}{\ell'_z} 
    \CG{1}{s'_z}{2}{\Delta\ell_z}{1}{s_z}\right)~,
  \end{aligned}
\end{equation}
with the coefficients (\ref{eq:1-5}).

\paragraph*{$\boldsymbol{|\Delta\ell|=2}$}
\label{sec:1-3}

In this case $S(j,\ell',\ell,1,1)$ can be non-vanishing only if
$\ell'=\ell+2$ and $j=\ell+1$, or if $\ell'=\ell-2$ and $j=\ell-1$~.
Using (\ref{eq:1-1}), (\ref{eq:redmat5}) and (\ref{eq:redmat3}), from
(\ref{eq:CGmaster}) and (\ref{eq:6j-2}) we obtain
\begin{equation}
  \label{eq:1-8}
  \begin{aligned}
    S(j,\ell_2,\ell,1,1) &= -C^{112}_{\ell_2\ell j}
    \langle\ell_2,\ell'_z|\veps{k_1k_2}(\Delta\ell_z)\verr{k_1}\verr{k_2}
    |\ell,\ell_z\rangle \,  
    \veps{h_1h_2}(\Delta\ell_z)^* \veps{h_1}(s'_z)^*\veps{h_2}(s_z)\\
    &=\sqrt{\frac{5}{3}} \sqrt{\frac{j(j+1)}{(2j+1)(2\ell_2+1)}}
    C^{112}_{\ell_2\ell j} \CG{\ell}{\ell_z}{2}{\dl_z}{\ell_2}{\ell'_z}
    \CG{1}{s'_z}{2}{\dl_z}{1}{s_z}~,\\
    C^{112}_{\ell_2\ell j} &= \frac{2j+1}{\sqrt{j(j+1)}}~.
  \end{aligned}
\end{equation}
The first line is the form needed for an addition theorem for
vector spherical harmonics.  Notice that the matrix element on that
line is traceless, because $\ell'\neq\ell$, and symmetric, so
we can replace $\veps{k_1k_2}(\Delta\ell_z)
\veps{h_1h_2}(\Delta\ell_z)^*$ there by
$\delta^{k_1h_1}\delta^{k_2h_2}$.  The second line is just (\ref{eq:6j-2}).

\subsection{$\boldsymbol{s=3/2=s'}$}
\label{sec:3h}

The spin matrix elements entering (\ref{eq:CGmaster}) when
$s=3/2=s'$ are given by
\begin{equation}
  \label{eq:3h-1}
  \begin{aligned}
  \veps{k}(\Delta\ell_z)^* \langle 3/2,s'_z | S^k | 3/2,s_z \rangle &= 
\frac{3}{2} \veps{k}(\Delta\ell_z)^*
\vchiU{A}{i}(s'_z)^*\sigma^k_{AB}\vchiU{B}{i}(s_z)~,\\
  \veps{ij}(\Delta\ell_z)^* \langle 3/2,s'_z | S^iS^j | 3/2,s_z \rangle &= 
-3 \veps{ij}(\Delta\ell_z)^*
\vchiU{A}{i}(s'_z)^*\vchiU{A}{j}(s_z)~,\\
  \veps{ijk}(\Delta\ell_z)^* \langle 3/2,s'_z | S^iS^jS^k | 3/2,s_z \rangle &= 
-\frac{3}{2} \veps{ijk}(\Delta\ell_z)^*
\vchiU{A}{i}(s'_z)^* \sigma^j_{AB} \vchiU{B}{k}(s_z)~.
  \end{aligned}
\end{equation}
The orbital angular momentum change is $0\le \Delta\ell \le 3$.

\paragraph*{$\boldsymbol{|\Delta\ell|=0}$}
\label{sec:3h-1}

Substituting (\ref{eq:3h-1}) in (\ref{eq:CGmaster}) we get
\begin{subequations}
  \label{eq:3h-2}
\begin{equation}
  \label{eq:3h-2a}
  \begin{aligned}
    S(j,\ell,\ell,3/2,3/2) &= C^{\frac{3}{2}\frac{3}{2}0}_{\ell\ell j} \delta_{\ell'_z
      \ell_z} \delta_{s'_zs_z} + \frac{3}{2}C^{\frac{3}{2}\frac{3}{2}1}_{\ell\ell
      j} \langle \ell,\ell'_z | \veps{i}(\dl_z )L^i | \ell,\ell_z
    \rangle \veps{k}(\dl_z)^* \vchiU{A}{h}(s'_z)^*
    \sigma^k_{AB} \vchiU{B}{h}(s_z)\\
    &\quad -3 C^{\frac{3}{2}\frac{3}{2}2}_{\ell\ell j} \langle \ell,\ell'_z |
    \veps{i_1i_2}(\dl_z)L^{i_1}L^{i_2} |\ell,\ell_z \rangle
    \veps{h_1h_2}(\dl_z)^* \vchiU{A}{h_1}(s'_z)^* \vchiU{A}{h_2}(s_z)\\
    &\quad -\frac{3}{2} C^{\frac{3}{2}\frac{3}{2}3}_{\ell\ell j} \langle \ell,\ell'_z |
    \veps{i_1 i_2 i_3}(\dl_z) L^{i_1}L^{i_2}L^{i_3} | \ell,\ell_z
    \rangle \veps{h_1h_2h_3}(\dl_z)^* \vchiU{A}{h_1}(s'_z)^*
    \sigma^{h_2}_{AB}\vchiU{B}{h_3}(s_z)~,
  \end{aligned}
\end{equation}  
where the coefficients $C^{\frac{3}{2}\frac{3}{2}\Delta}_{\ell\ell j}$
given in (\ref{eq:CGmaster}) in this case take the form
\begin{equation}
  \label{eq:3h-2b}
  \begin{aligned}
C^{\frac{3}{2}\frac{3}{2}0}_{\ell\ell j} &= \frac{1}{4}\,\frac{2j+1}{2\ell+1}~,\\
C^{\frac{3}{2}\frac{3}{2}1}_{\ell\ell j} &= \frac{2}{5}\, \frac{j-\ell}{2\ell+1}\,
\left(j+\frac{1}{2}+4 \left(j-\ell-\frac{3}{2}\right) (j-\ell)
  \left(j-\ell+\frac{3}{2}\right)
\right)\frac{1}{\widetilde{D}_{\ell j}}~,\\ 
C^{\frac{3}{2}\frac{3}{2}2}_{\ell\ell j} &= -\frac{1}{2}\, \frac{1}{2\ell+1}\, \left(
  \frac{5}{4} - (j-\ell)^2 \right) \left( j+\frac{1}{2}-2(j-\ell)^3 +
  \frac{9}{2} (j-\ell) \right) \frac{1}{D_{\ell j}}~,\\
C^{\frac{3}{2}\frac{3}{2}3}_{\ell\ell j} &= -\frac{1}{2}\,
\frac{(-1)^{j-\ell-1/2}}{|j-\ell|(2\ell+1)}\, \frac{1}{D_{\ell j}}~,\\  
\widetilde{D}_{\ell j} &= \left( \ell+1+\frac{2}{3}\left(j-\ell-\frac{3}{2}\right)
  \left(j-\ell+\frac{1}{2}\right)(j-\ell+1) \right)~,\\
D_{\ell j} &= \widetilde{D}_{\ell j}\times
\left( j - \frac{2}{3}\left(j-\ell-\frac{3}{2}\right)
\left(j-\ell+\frac{1}{2}\right)(j-\ell+1) \right)~.
  \end{aligned}
\end{equation}
\end{subequations}
As above, we notice that the matrix elements in (\ref{eq:3h-2a}) can
also be written as,
\begin{equation}
  \label{eq:3halfaux1}
 \begin{aligned}
  \veps{p}(\Delta\ell_z) \langle\ell,\ell'_z|L^p|\ell,\ell_z\rangle
  \veps{k}(\Delta\ell_z)^*
  \vchiU{A}{i}(s'_z)^*\sigma^k_{AB}\vchiU{B}{i}(s_z)&=
\langle \ell,\ell'_z | L^k |
  \ell,\ell_z \rangle
  \vchiU{A}{i}(s'_z)^*\sigma^k_{AB}\vchiU{B}{i}(s_z)~, \\ 
  2\veps{pq}(\Delta\ell_z)
  \langle\ell,\ell'_z|L^pL^q|\ell,\ell_z\rangle
  \veps{ij}(\Delta\ell_z)^* \vchiU{A}{i}(s'_z)^* \vchiU{B}{j}(s_z) &=
  \langle\ell,\ell'_z|L^{\{i}L^{j\}_0}|\ell,\ell_z\rangle
  \vchiU{A}{i}(s'_z)^* \vchiU{A}{j}(s_z) ~,\\
6\veps{pqr}(\Delta\ell_z)\langle\ell,\ell'_z|L^pL^qL^r|\ell,\ell_z\rangle
\veps{ijk}(\Delta\ell_z)^* \vchiU{A}{i}(s'_z)^*
\sigma^k_{AB}\vchiU{B}{j}(s_z) 
&=
\langle \ell,\ell'_z | L^{\{i}L^{j}L^{k\}_0} |\ell,\ell_z \rangle
\\
&\qquad\times \vchiU{A}{i}(s'_z)^* \sigma^j_{AB}\vchiU{B}{k}(s_z) ~.
 \end{aligned}
\end{equation}
With the matrix elements written as in (\ref{eq:3h-2a}), 
$S(j,\ell,\ell,3/2,3/2)$ takes the form needed in the
derivation of addition theorems for spin-3/2 spherical harmonics.

Similarly, (\ref{eq:6j-2}) can be rewritten as
\begin{equation}
\label{eq:3h-3}
  \begin{gathered}
  \begin{aligned}
    S(j,\ell,\ell,3/2,3/2) &= \kappa^{\frac{3}{2}\frac{3}{2}0}_{\ell\ell j} \delta_{\ell'_z \ell_z}
    \delta_{s'_zs_z} + \kappa^{\frac{3}{2}\frac{3}{2}1}_{\ell\ell j} \CG{\ell}{\ell_z}{1}{\Delta\ell_z}{\ell}{\ell'_z}
    \CG{3/2}{s'_z}{1}{\Delta\ell_z}{3/2}{s_z}\\
    &\quad +\kappa^{\frac{3}{2}\frac{3}{2}2}_{\ell\ell j}
    \CG{\ell}{\ell_z}{2}{\Delta\ell_z}{\ell}{\ell'_z}
    \CG{3/2}{s'_z}{2}{\Delta\ell_z}{3/2}{s_z}\\
    &\quad +\kappa^{\frac{3}{2}\frac{3}{2}3}_{\ell\ell j}\CG{\ell}{\ell_z}{3}{\Delta\ell_z}{\ell}{\ell'_z}
    \CG{3/2}{s'_z}{3}{\Delta\ell_z}{3/2}{s_z}~,
  \end{aligned}\\
    \kappa^{\frac{3}{2}\frac{3}{2}0}_{\ell\ell j} = C^{\frac{3}{2}\frac{3}{2}0}_{\ell\ell j},
\quad
    \kappa^{\frac{3}{2}\frac{3}{2}1}_{\ell\ell j} =  \sqrt{\frac{15}{4}} \sqrt{\ell(\ell+1)}
    C^{\frac{3}{2}\frac{3}{2}1}_{\ell\ell j},
\quad
    \kappa^{\frac{3}{2}\frac{3}{2}2}_{\ell\ell j} = \sqrt{\frac{5}{4}} 
    \sqrt{\ell(\ell+1)} \sqrt{(2\ell-1)(2\ell+3)}
    C^{\frac{3}{2}\frac{3}{2}2}_{\ell\ell j},\\
    \kappa^{\frac{3}{2}\frac{3}{2}3}_{\ell\ell j} = \frac{3}{2} \sqrt{\frac{7}{20}}
    \sqrt{(\ell-1)\ell(\ell+1)(\ell+2)} \sqrt{(2\ell-1)(2\ell+3)} 
    C^{\frac{3}{2}\frac{3}{2}3}_{\ell\ell j}~,
  \end{gathered}
\end{equation}
Eqs.\ (\ref{eq:3h-2}) and (\ref{eq:3h-3}) are of course related by
the WE theorem, with the reduced matrix elements in appendix
\ref{sec:matele}. 

\paragraph*{$\boldsymbol{|\Delta\ell|=1}$}
\label{sec:3h-2}

In this case $S(j,\ell',\ell,3/2,3/2)$ can be non-vanishing only if 
$\ell'=\ell+1$ and $j=\ell+3/2$, $\ell+1/2$, $\ell-1/2$, or if
$\ell'=\ell-1$ and $j=\ell+1/2$, $\ell-1/2$, $\ell-3/2$.
Evaluating the spin matrix elements with (\ref{eq:3h-1}) we find,
\begin{subequations}
  \label{eq:3h-4}
\begin{equation}
  \label{eq:3h-4a}
  \begin{aligned}
    S(j,\ell_1,\ell,3/2,3/2) &=
    \frac{3}{2}C^{\frac{3}{2}\frac{3}{2}1}_{\ell_1\ell j}
    \langle\ell_1,\ell'_z | \veps{k}(\dl_z) \verr{k} |
    \ell,\ell_z\rangle \, \veps{h}(\dl_z)^*
    \vchiU{A}{j}(s'_z)^*\sigma^h_{AB}\vchiU{B}{j}(s_z) \\
    &-3 C^{\frac{3}{2}\frac{3}{2}2}_{\ell_1\ell j}
    \langle\ell_1,\ell'_z | \veps{i_1i_2}(\dl_z)\verr{i_1}L^{i_2} |
    \ell,\ell_z\rangle \, \veps{j_1j_2}(\dl_z)^*
    \vchiU{A}{j_1}(s'_z)^*
    \vchiU{A}{j_2}(s_z)\\
    &-\frac{3}{2}C^{\frac{3}{2}\frac{3}{2}3}_{\ell_1\ell j}
    \langle\ell_1,\ell'_z | \veps{i_1i_2i_3}(\dl_z) \verr{i_1} L^{i_2}
    L^{i_3}|\ell,\ell_z\rangle \, \veps{j_1j_2j_3}(\dl_z)^*
    \vchiU{A}{j_1}(s'_z)^*\sigma^{j_2}_{AB}\vchiU{B}{j_3}(s_z)~,
  \end{aligned}
\end{equation}
with the coefficients $C^{\frac{3}{2}\frac{3}{2}\Delta}_{\ell_1\ell j}$ given by
\begin{align}
C^{\frac{3}{2}\frac{3}{2}1}_{\ell_1\ell j} &= -\frac{1}{5}\,
 \frac{j+1/2}{2\lave+1}
\frac{\sqrt{\left(\rule{0pt}{9pt} j+\lave+1\right)^2 - 4}\, 
\sqrt{4 - \left(\rule{0pt}{9pt} j-\lave\right)^2}}
{\sqrt{\lave(\lave+1)}}
~,\nonumber\\
C^{\frac{3}{2}\frac{3}{2}2}_{\ell_1\ell j} &= -2 \frac{\left(\rule{0pt}{10pt}
    3\lave-j+1\right)}
{(2\lave+1)\sqrt{\lave(\lave+1)}}\,
\frac{ \left(\rule{0pt}{10pt}
    (j-\lave)(j+1/2)+1/2\right) \left(\rule{0pt}{10pt}
    5(j-\lave)^2-4\right)} {\sqrt{\left(
      j+\lave+1\right)^2 - 4} \sqrt{4 - \left(
      j-\lave\right)^2}}~,  \label{eq:3h-4b}\\
C^{\frac{3}{2}\frac{3}{2}3}_{\ell_1\ell j} &=-4
\frac{\sqrt{\lave(\lave+1)}}{(2\lave+1)j(j+1)} 
\frac{\left(\rule{0pt}{10pt} 3 j-\lave+1\right)
\left(\rule{0pt}{10pt} 3/2(j-\lave)^2 -1\right)}
{\sqrt{\left( j+\lave+1\right)^2 - 4} \sqrt{4 - \left(
      j-\lave\right)^2}}~.\nonumber
\end{align}
\end{subequations}
This expression in terms of matrix elements of irreducible tensor
operators is needed in the derivation of addition theorems for
spin-3/2 spherical harmonics.  We notice also that the matrix elements
in (\ref{eq:3h-4a}) can be written without standard tensors as,
\begin{equation}
  \label{eq:3h-4x}
  \begin{gathered}
    \langle\ell_1,\ell'_z | \veps{k}(\dl_z) \verr{k} |
    \ell,\ell_z\rangle \, \veps{h}(\dl_z)^*
    \vchiU{A}{j}(s'_z)^*\sigma^h_{AB}\vchiU{B}{j}(s_z) =
    \langle\ell_1,\ell'_z | \verr{k} | \ell,\ell_z\rangle \,
    \vchiU{A}{j}(s'_z)^*\sigma^k_{AB}\vchiU{B}{j}(s_z)~,  \\
    \langle\ell_1,\ell'_z | \veps{i_1i_2}(\dl_z)\verr{i_1}L^{i_2} |
    \ell,\ell_z\rangle \, \veps{j_1j_2}(\dl_z)^*
    \vchiU{A}{j_1}(s'_z)^* \vchiU{A}{j_2}(s_z) = \frac{1}{2}
    \langle\ell_1,\ell'_z | \verr{\{i_1}L^{i_2\}} | \ell,\ell_z\rangle
    \, \vchiU{A}{i_1}(s'_z)^* \vchiU{A}{i_2}(s_z)~,
    \\
    \begin{aligned}
    \langle\ell_1,\ell'_z | \veps{i_1i_2i_3}(\dl_z) \verr{i_1} L^{i_2}
    L^{i_3}|\ell,\ell_z\rangle \, \veps{j_1j_2j_3}(\dl_z)^*
    \vchiU{A}{j_1}(s'_z)^*\sigma^{j_2}_{AB}\vchiU{B}{j_3}(s_z) &=
    \frac{1}{6} \langle\ell_1,\ell'_z | \verr{\{i_1} L^{i_2}
    L^{i_3\}_0}|\ell,\ell_z\rangle \\
    &\quad\times
    \vchiU{A}{i_1}(s'_z)^*\sigma^{i_2}_{AB}\vchiU{B}{i_3}(s_z)~. 
    \end{aligned}
  \end{gathered}
\end{equation}
Analogously, (\ref{eq:6j-2}) can be
written as
\begin{subequations}
\label{eq:3h-5}
\begin{equation}
  \label{eq:3h-5a}
  \begin{aligned}
S(j,\ell_1,\ell,3/2,3/2) &= \kappa^{\frac{3}{2}\frac{3}{2}1}_{\ell_1\ell j} 
\CG{\ell}{\ell_z}{1}{\Delta\ell_z}{\ell_1}{\ell'_z}  
\CG{3/2}{s'_z}{1}{\Delta\ell_z}{3/2}{s_z}  \\
&\quad+
\kappa^{\frac{3}{2}\frac{3}{2}2}_{\ell_1\ell j}
\CG{\ell}{\ell_z}{2}{\Delta\ell_z}{\ell_1}{\ell'_z}  
\CG{3/2}{s'_z}{2}{\Delta\ell_z}{3/2}{s_z}\\
&\quad+
\kappa^{\frac{3}{2}\frac{3}{2}3}_{\ell_1\ell j}
\CG{\ell}{\ell_z}{3}{\Delta\ell_z}{\ell_1}{\ell'_z}  
\CG{3/2}{s'_z}{3}{\Delta\ell_z}{3/2}{s_z}~,
  \end{aligned}
\end{equation}
with,
\begin{equation}
  \label{eq:3h-5b}
\begin{aligned}
  \kappa^{\frac{3}{2}\frac{3}{2}1}_{\ell_1\ell j} &= \frac{1}{2}
  \sqrt{\frac{15}{2}} \frac{\dl}{\sqrt{2\ell_1+1}} \sqrt{2\lave+1}
  \, C^{\frac{3}{2}\frac{3}{2}1}_{\ell_1\ell j}~,\\
\kappa^{\frac{3}{2}\frac{3}{2}2}_{\ell_1\ell j} &= \frac{1}{4}
\sqrt{\frac{15}{2}} \frac{\dl}{\sqrt{2\ell_1+1}}
\sqrt{(2\lave-1) (2\lave+1)(2\lave+3)}\,
C^{\frac{3}{2}\frac{3}{2}2}_{\ell_1\ell j}~, \\  
\kappa^{\frac{3}{2}\frac{3}{2}3}_{\ell_1\ell j} &= \frac{1}{8}
\sqrt{\frac{21}{5}} \frac{\dl}{\sqrt{2\ell_1+1}} 
\sqrt{(2\lave-2)(2\lave-1)(2\lave+1)(2\lave+3)(2\lave+4)}\,  
C^{\frac{3}{2}\frac{3}{2}3}_{\ell_1\ell j}~,
\end{aligned}
\end{equation}
\end{subequations}
where the coefficients $C^{\frac{3}{2}\frac{3}{2}\Delta}_{\ell_1\ell
  j}$ are given in (\ref{eq:3h-4b}).

\paragraph*{$\boldsymbol{|\Delta\ell|=2}$}
\label{sec:3h-3}

In this case $S(j,\ell',\ell,3/2)$ can be non-vanishing only if 
$\ell'=\ell+2$ and $j=\ell+3/2,\ell+1/2$ or $\ell'=\ell-2$ and
$j=\ell-1/2,\ell-3/2$. Taking into account this reduced set of $j$
values, and the spin matrix elements (\ref{eq:3h-1}), eq.\
(\ref{eq:CGmaster}) takes the form
\begin{equation}
  \label{eq:3h-6}
  \begin{gathered}
    \begin{aligned}
      S(j,\ell_2,\ell,3/2,3/2) &= -3
      C^{\frac{3}{2}\frac{3}{2}2}_{\ell_2\ell j}
      \langle\ell_2,\ell'_z|\veps{k_1
        k_2}(\dl_z)\verr{k_1}\verr{k_2}|\ell, \ell_z\rangle
      \veps{h_1h_2}(\dl_z)^* \vchiU{A}{h_1}(s'_z)^*
      \vchiU{A}{h_2}(s_z) \\
      &- \frac{3}{2} C^{\frac{3}{2}\frac{3}{2}3}_{\ell_2\ell j}
      \langle\ell_2,\ell'_z|\veps{k_1k_2k_3}(\dl_z)\verr{k_1}\verr{k_2}L^{k_3}|\ell,
      \ell_z\rangle \veps{h_1h_2h_3}(\dl_z)^* \vchiU{A}{h_1}(s'_z)^*
      \sigma^{h_2}_{AB}\vchiU{B}{h_3}(s_z)~,
    \end{aligned}\\
C^{\frac{3}{2}\frac{3}{2}2}_{\ell_2\ell j} = \frac{2}{\sqrt{3}}
\frac{3j-2\lave+1/2}{\sqrt{(2j-1)(2j+3)}}~,
\quad
C^{\frac{3}{2}\frac{3}{2}3}_{\ell_2\ell j} = \frac{8}{\sqrt{3}}
\frac{j-\lave}{\sqrt{(2j-1)(2j+3)}}~.
  \end{gathered}
\end{equation}
We note here that the matrix elements appearing in this equation can
also be written as
\begin{equation}
  \label{eq:3h-6aux}
  \begin{gathered}
  \langle\ell',\ell'_z|\veps{k_1
        k_2}(\dl_z)\verr{k_1}\verr{k_2}|\ell, \ell_z\rangle
      \veps{h_1h_2}(\dl_z)^* \vchiU{A}{h_1}(s'_z)^*
      \vchiU{A}{h_2}(s_z) =
  \langle\ell',\ell'_z|\verr{i_1}\verr{i_2}|\ell, \ell_z\rangle
      \vchiU{A}{i_1}(s'_z)^* \vchiU{A}{i_2}(s_z)~,\\
\begin{aligned}
\lefteqn{
      \langle\ell',\ell'_z|\veps{k_1k_2k_3}(\dl_z)\verr{k_1}\verr{k_2}L^{k_3}|\ell, 
      \ell_z\rangle \veps{h_1h_2h_3}(\dl_z)^* \vchiU{A}{h_1}(s'_z)^*
      \sigma^{h_2}_{AB}\vchiU{B}{h_3}(s_z) =
      }\hspace{48ex}\\
      & \frac{1}{6}
      \langle\ell',\ell'_z|\verr{\{i_1}\verr{i_2}L^{i_3\}_0}|\ell, 
      \ell_z\rangle  \vchiU{A}{i_1}(s'_z)^*
      \sigma^{i_2}_{AB}\vchiU{B}{i_3}(s_z)~,
      \end{aligned}
  \end{gathered}
\end{equation}
where the matrix element in the r.h.s.\ of the first equality is
traceless because $\ell'\neq \ell$.   Similarly, from eq.\
(\ref{eq:6j-2}) we get 
\begin{equation}
  \label{eq:3h-7}
  \begin{aligned}
S(j,\ell_2,\ell,3/2,3/2) &= 
\kappa^{\frac{3}{2}\frac{3}{2}2}_{\ell_2\ell j}
\CG{\ell}{\ell_z}{2}{\Delta\ell_z}{\ell_2}{\ell'_z}  
\CG{3/2}{s'_z}{2}{\Delta\ell_z}{3/2}{s_z}\\
&\quad+
\kappa^{\frac{3}{2}\frac{3}{2}3}_{\ell_2\ell j}
\CG{\ell}{\ell_z}{3}{\Delta\ell_z}{\ell_2}{\ell'_z}  
\CG{3/2}{s'_z}{3}{\Delta\ell_z}{3/2}{s_z}~,\\
%
%
  \kappa^{\frac{3}{2}\frac{3}{2}2}_{\ell_2\ell j} &=
  \sqrt{\frac{15}{2}}\sqrt{2\ell+1} \sqrt{\frac{\lave (\lave+1)}
    {(2\lave-1)(2\lave+1)(2\lave+3)}}
  C^{\frac{3}{2}\frac{3}{2}2}_{\ell_2\ell j} ~,\\
\quad
  \kappa^{\frac{3}{2}\frac{3}{2}3}_{\ell_2\ell j} &= \frac{1}{2}
  \sqrt{\frac{21}{2}} \sqrt{2\ell+1} 
  \sqrt{\frac{(\lave-1)\lave (\lave+1)(\lave+2)}
    {(2\lave-1)(2\lave+1)(2\lave+3)}}
  C^{\frac{3}{2}\frac{3}{2}3}_{\ell_2\ell j} ~,
  \end{aligned}
\end{equation}
with the coefficients $C^{\frac{3}{2}\frac{3}{2}\Delta}_{\ell_2\ell
  j}$ of (\ref{eq:3h-6}).

\paragraph*{$\boldsymbol{|\Delta\ell|=3}$}
\label{sec:3h-4}

In this case $S(j,\ell',\ell,3/2,3/2)$ can be non-vanishing only if 
$\ell'=\ell+3$ and $j=\ell+3/2$, or $\ell'=\ell-3$ and $j=\ell-3/2$.
Eq.\ (\ref{eq:CGmaster}) then reduces to
\begin{equation}
  \label{eq:3h-8}
  \begin{aligned}
  S(j,\ell_3,\ell,3/2,3/2) &=
  -\frac{3}{2}C^{\frac{3}{2}\frac{3}{2}3}_{\ell_3\ell j}
   \langle\ell_3,\ell'_z|\veps{k_1k_2k_3}(\dl_z)\verr{k_1}\verr{k_2}\verr{k_3}|\ell,
   \ell_z\rangle \\
   &\quad \times\veps{h_1h_2h_3}(\dl_z)^* \vchiU{A}{h_1}(s'_z)^*
   \sigma^{h_2}_{AB}\vchiU{B}{h_3}(s_z)~,  \\
  C^{\frac{3}{2}\frac{3}{2}3}_{\ell_3\ell j} &= -\frac{8}{3}
  \sqrt{\frac{j(j+1)}{(2j-1)(2j+3)}}~, 
  \end{aligned}
\end{equation}
which is the form needed to formulate an addition theorem for
spin-3/2 spherical harmonics with $\ell'=\ell\pm 3$. The matrix
element in (\ref{eq:3h-8}) can also be written 
\begin{equation}
  \label{eq:3h-9}
  \begin{split}
   \langle\ell',\ell'_z|\veps{k_1k_2k_3}(\dl_z)\verr{k_1}\verr{k_2}\verr{k_3}|\ell,
   \ell_z\rangle 
    \veps{h_1h_2h_3}(\dl_z)^* \vchiU{A}{h_1}(s'_z)^*
    \sigma^{h_2}_{AB}\vchiU{B}{h_3}(s_z) =
    \hspace{25ex}\\
    =\frac{1}{6}
   \langle\ell',\ell'_z|\verr{\{k_1}\verr{k_2}\verr{k_3\}_0}|\ell,
   \ell_z\rangle 
    \vchiU{A}{k_1}(s'_z)^* \sigma^{k_2}_{AB}\vchiU{B}{k_3}(s_z)~. 
  \end{split}
\end{equation}
Similarly, (\ref{eq:6j-2}) reduces to
\begin{equation}
  \label{eq:3h-a}
  \begin{aligned}
  S(j,\ell_3,\ell,3/2,3/2) &=
\kappa^{\frac{3}{2}\frac{3}{2}3}_{\ell_3\ell j} 
  \CG{\ell}{\ell_z}{3}{\Delta\ell_z}{\ell_3}{\ell'_z}  
  \CG{3/2}{s'_z}{3}{\Delta\ell_z}{3/2}{s_z}~,\\
  \kappa^{\frac{3}{2}\frac{3}{2}3}_{\ell_3\ell j} &=
  \frac{\sqrt{7}}{32} \dl \sqrt{\frac{2\ell+1}{\lave+2}}
  \sqrt{\frac{(2\lave-1) (2\lave+1) (2\lave+3)}{(\lave-1) \lave
      (\lave+1)}}C^{\frac{3}{2}\frac{3}{2}3}_{\ell_3\ell j} ~,
  \end{aligned}
\end{equation}
with $C^{\frac{3}{2}\frac{3}{2}3}_{\ell_3\ell j}$ from (\ref{eq:3h-8}).

\subsection{The spin transition operator} 
\label{sec:spintrans}

We define the spin transition operator $\vec{T}$ as a spin vector
operator, commuting with all orbital operators, and satisfying
\begin{equation}
  \label{eq:Top}
  [S^i,T^j] = i\varepsilon^{ijk} T^k~,
  \qquad
  \langle s' || \veps{i}T^i || s \rangle = \frac{s'-s}{\sqrt{2}}
  \sqrt{\frac{2\save+1}{2s'+1}} \left(\delta_{s'(s+1)}+\delta_{s'(s-1)}\right).
\end{equation}
From (\ref{eq:Top}) we have,
\begin{equation}
  \label{eq:Top2}
  \langle s',s'_z | \vec{T}^2 | s, s_z\rangle = 
  \begin{cases}
    \delta_{s's}\delta_{s'_zs_z}& \text{if $s'>1/2$ or $s>1/2$}\\
    \frac{3}{4}\delta_{s'_zs_z}& \text{if $s'=s=1/2$}\
  \end{cases}~.
\end{equation}
 From (\ref{eq:Top}) and (\ref{eq:Top2}) the operator $\vec{T}$ is seen
to be the spin-space analog of the orbital operator $\verr{}$.  In
particular, the reduced matrix elements of tensor products of
$\verr{}$ and $\vec{L}$ given in Appendix \ref{sec:matele} apply
without changes to tensor products of $\vec{T}$ and $\vec{S}$.

\subsection{$\boldsymbol{s',s=1,0}$ or $\boldsymbol{0,1}$}  
\label{sec:10}

The relevant spin matrix element in this
case is,
\begin{equation}
  \label{eq:10-1}
  \langle 1, s'_z | T^h | 0, 0 \rangle = \frac{1}{\sqrt{3}}
  \veps{h}(s'_z)^*~. 
\end{equation}
$|\Delta\ell|$ can take the values 0 and 1.

\paragraph*{$\boldsymbol{|\Delta\ell|=0}$}
\label{sec:10-1}

$S(j,\ell,\ell,1,0)$ (resp.\ $S(j,\ell,\ell,0,1)$) can be non-vanishing only if
$j=\ell$ (resp.\ $j=\ell'$). 
Then, (\ref{eq:CGmaster}) and (\ref{eq:6j-2}) take the form,
\begin{equation}
  \label{eq:10-2}
  \begin{gathered}
  S(\ell,\ell,\ell,1,0) = \frac{1}{\sqrt{3}} C^{101}_{\ell\ell \ell} \langle
  \ell, \ell'_z | L^k | \ell, \ell_z \rangle \veps{k}(s'_z)^*
  = -\sqrt{3} \CG{\ell}{\ell_z}{1}{\dl_z}{\ell}{\ell'_z}
  \CG{1}{s'_z}{1}{\dl_z}{0}{0}~,\\
  C^{101}_{\ell\ell \ell} = \sqrt{\frac{3}{\ell(\ell+1)}}~.
  \end{gathered}
\end{equation}
Similarly, 
\begin{equation}
  \label{eq:10-3}
  S(\ell,\ell,\ell,0,1) = \frac{1}{\sqrt{3}} C^{011}_{\ell\ell \ell} \langle
  \ell, \ell'_z | L^k | \ell, \ell_z \rangle \veps{k}(s_z)
  = \CG{\ell}{\ell_z}{1}{\dl_z}{\ell}{\ell'_z}
  \CG{0}{0}{1}{\dl_z}{1}{s_z}~,
\end{equation}
with $C^{011}_{\ell\ell \ell} = C^{101}_{\ell\ell \ell}$.

\paragraph*{$\boldsymbol{|\Delta\ell|=1}$}
\label{sec:10-2}

In this case $S(j,\ell',\ell,s',s)$ can be non-vanishing only if
$j=\ell$ (if $s=0$) or $\ell'$ (if $s'=0$).  Thus, with the spin
matrix element (\ref{eq:10-1}) we get, from (\ref{eq:CGmaster}) and
(\ref{eq:6j-2})~, 
\begin{equation}
  \label{eq:10-4}
  \begin{gathered}
  S(\ell,\ell_1,\ell,1,0) = \frac{C^{101}_{\ell_1\ell\ell}}{\sqrt{3}}
  \langle \ell_1,\ell'_z | \verr{k} | \ell,\ell_z\rangle 
  \veps{k}(s'_z)^*
  = \sqrt{3} \sqrt{\frac{2\ell+1}{2\ell_1+1}}
  \CG{\ell}{\ell_z}{1}{\dl_z}{\ell_1}{\ell'_z}
  \CG{1}{s'_z}{1}{\dl_z}{0}{0}~,\\
  C^{101}_{\ell_1\ell\ell} = -\sqrt{6} \dl
  \sqrt{\frac{2\ell+1}{2\lave+1}}~. 
  \end{gathered}
\end{equation}
Analogously,
\begin{equation}
  \label{eq:10-5}
  \begin{gathered}
  S(\ell_1,\ell_1,\ell,0,1) = \frac{C^{011}_{\ell_1\ell\ell_1}}{\sqrt{3}}
  \langle \ell_1,\ell'_z | \verr{k} | \ell,\ell_z\rangle 
  \veps{k}(s'_z)
  = 
  \CG{\ell}{\ell_z}{1}{\dl_z}{\ell_1}{\ell'_z}
  \CG{1}{s'_z}{1}{\dl_z}{0}{0}~,\\
  C^{011}_{\ell_1\ell\ell'} = \sqrt{6} \dl
  \sqrt{\frac{2\ell'+1}{2\lave+1}}~. 
  \end{gathered}  
\end{equation}

\subsection{$\boldsymbol{s',s=3/2,1/2}$ or
  $\boldsymbol{1/2,3/2}$} 
\label{sec:3h1h}

The spin matrix elements
appearing in (\ref{eq:CGmaster}) in this case are
\begin{equation}
  \label{eq:3h1h-1}
  \begin{aligned}
  \langle 3/2,s'_z | \veps{h}(\dl_z)^* T^h | 1/2, s_z \rangle &=
  \frac{1}{2} \sqrt{\frac{3}{2}} \veps{h}(\dl_z)^*
  \vchiU{A}{h}(s'_z)^* \vchi{A}(s_z)~,\\
  \langle 3/2,s'_z | \veps{h_1h_2}(\dl_z)^* T^{h_1}S^{h_2} | 1/2, s_z
  \rangle &=  \frac{1}{4} \sqrt{\frac{3}{2}} \veps{h_1h_2}(\dl_z)^*
  \vchiU{A}{h_1}(s'_z)^* \sigma^{h_2}_{AB} \vchi{B}(s_z)~.
  \end{aligned}
\end{equation}
There are three possible values for $0\le |\Delta\ell| \le 2$.

\paragraph*{$\boldsymbol{|\Delta\ell|=0}$}
\label{sec:3h1h-1}

If $s'=3/2$, $s=1/2$, and $\ell'=\ell$, then $S(j,\ell,\ell,3/2,1/2)$
can be non-vanishing only if $j=\ell\pm1/2$.  In the case $s'=1/2$,
$s=3/2$, it must be $j=\ell'\pm1/2$.  
Substituting (\ref{eq:3h1h-1}) in (\ref{eq:CGmaster}) we obtain
\begin{subequations}
  \label{eq:3h1h-2}
\begin{equation}
  \label{eq:3h1h-2a}
  \begin{aligned}
  S(j,\ell,\ell,3/2,1/2) &= \frac{1}{2} \sqrt{\frac{3}{2}}
  C^{\frac{3}{2}\frac{1}{2}1}_{\ell\ell j} \langle \ell, \ell'_z | \veps{k}(\dl_z)
  L^k | \ell, \ell_z \rangle \veps{h}(\dl_z)^* \vchiU{A}{h}(s'_z)^*
  \vchi{A}(s_z) \\ 
  &\quad + 
  \frac{1}{4} \sqrt{\frac{3}{2}} C^{\frac{3}{2}\frac{1}{2}2}_{\ell\ell
    j} \langle \ell, \ell'_z | \veps{k_1 k_2}(\dl_z) L^{k_1} L^{k_2} |
  \ell, \ell_z \rangle  \veps{h_1h_2}(\dl_z)^*
  \vchiU{A}{h_1}(s'_z)^* \sigma^{h_2}_{AB} \vchi{B}(s_z)~,
  \end{aligned}
\end{equation}
with
\begin{equation}
  \label{eq:3h1h-2b}
C^{\frac{3}{2}\frac{1}{2}1}_{\ell\ell j} = \frac{2}{2\ell+1}
\sqrt{\frac{2j-\ell+1/2}{2\ell-j+1/2}}~,
\qquad
C^{\frac{3}{2}\frac{1}{2}2}_{\ell\ell j} = 8 \frac{j-\ell}{2\ell+1} 
\frac{1}{\sqrt{(2j-\ell+1/2)(2\ell-j+1/2)}}~.
\end{equation}
\end{subequations}
The matrix elements in (\ref{eq:3h1h-2a}) can also be written without
standard tensors as,
\begin{equation}
  \label{eq:3h1h-3}
  \begin{gathered}
    \langle \ell, \ell'_z | \veps{k}(\dl_z) L^k | \ell, \ell_z \rangle
    \veps{h}(\dl_z)^* \vchiU{A}{h}(s'_z)^* \vchi{A}(s_z) = \langle
    \ell, \ell'_z | L^k | \ell, \ell_z \rangle \vchiU{A}{k}(s'_z)^*
    \vchi{A}(s_z) ~,\\
    \begin{aligned}
      \lefteqn{
    \langle \ell, \ell'_z | \veps{k_1 k_2}(\dl_z) L^{k_1} L^{k_2} |
    \ell, \ell_z \rangle \veps{h_1h_2}(\dl_z)^* \vchiU{A}{h_1}(s'_z)^*
    \sigma^{h_2}_{AB} \vchi{B}(s_z) = }\hspace{53ex}\\
    & =\frac{1}{2}\langle \ell, \ell'_z |
    L^{\{k_1} L^{k_2\}_0} | \ell, \ell_z \rangle
     \vchiU{A}{k_1}(s'_z)^* \sigma^{k_2}_{AB}
    \vchi{B}(s_z)~.
    \end{aligned}
  \end{gathered}
\end{equation}
Similarly, (\ref{eq:6j-2}) takes the simpler form,
\begin{equation}
  \label{eq:3h1h-4}
  \begin{split}
  S(j,\ell,\ell,3/2,1/2) = -\frac{\sqrt{3}}{2} \sqrt{\ell(\ell+1)}
  C^{\frac{3}{2}\frac{1}{2}1}_{\ell\ell j}
  \CG{\ell}{\ell_z}{1}{\dl_z}{\ell}{\ell'_z}
  \CG{3/2}{s'_z}{1}{\dl_z}{1/2}{s_z}
  \hspace{20ex}\\ 
  - \frac{\sqrt{5}}{8} \sqrt{\ell(\ell+1)} \sqrt{(2\ell-1)(2\ell+3)}
  C^{\frac{3}{2}\frac{1}{2}2}_{\ell\ell j}
  \CG{\ell}{\ell_z}{2}{\dl_z}{\ell}{\ell'_z}
  \CG{3/2}{s'_z}{2}{\dl_z}{1/2}{s_z}~,
  \end{split}  
\end{equation}
with the coefficients given in  (\ref{eq:3h1h-2b}).

The case $s'=1/2$, $s=3/2$ is completely analogous. From
(\ref{eq:CGmaster}) we find
\begin{equation}
  \label{eq:3h1h-5}
  \begin{aligned}
  S(j,\ell,\ell,1/2,3/2) &= \frac{1}{2} \sqrt{\frac{3}{2}}
  C^{\frac{1}{2}\frac{3}{2}1}_{\ell\ell j} \langle \ell, \ell'_z | \veps{k}(\dl_z)
  L^k | \ell, \ell_z \rangle \veps{h}(\dl_z)^* \vchi{A}(s'_z)^*
  \vchiU{A}{h}(s_z) \\ 
  &\quad + 
  \frac{1}{4} \sqrt{\frac{3}{2}} C^{\frac{1}{2}\frac{3}{2}2}_{\ell\ell
    j} \langle \ell, \ell'_z | \veps{k_1 k_2}(\dl_z) L^{k_1} L^{k_2} |
  \ell, \ell_z \rangle  \veps{h_1h_2}(\dl_z)^*
  \vchi{A}(s'_z)^* \sigma^{h_2}_{AB} \vchiU{B}{h_1}(s_z)~,
  \end{aligned}
\end{equation}
with $C^{\frac{1}{2}\frac{3}{2}\Delta}_{\ell\ell j} =
C^{\frac{3}{2}\frac{1}{2}\Delta}_{\ell\ell j}$ (see
(\ref{eq:3h1h-2b})), and from 
(\ref{eq:6j-2})
\begin{equation}
  \label{eq:3h1h-6}
  \begin{split}
  S(j,\ell,\ell,1/2,3/2) = \frac{1}{2} \sqrt{\frac{3}{2}}
  \sqrt{\ell(\ell+1)} C^{\frac{1}{2}\frac{3}{2}1}_{\ell\ell j}
  \CG{\ell}{\ell_z}{1}{\dl_z}{\ell}{\ell'_z}
  \CG{1/2}{s'_z}{1}{\dl_z}{3/2}{s_z}
  \hspace{15ex}\\ 
  + \frac{1}{8} \sqrt{\frac{5}{2}} \sqrt{\ell(\ell+1)}
  \sqrt{(2\ell-1)(2\ell+3)} 
  C^{\frac{1}{2}\frac{3}{2}2}_{\ell\ell j}
  \CG{\ell}{\ell_z}{2}{\dl_z}{\ell}{\ell'_z}
  \CG{1/2}{s'_z}{2}{\dl_z}{3/2}{s_z}~.
  \end{split}  
\end{equation} 
The expressions (\ref{eq:3h1h-2}) and (\ref{eq:3h1h-5}) will be used
below to derive addition theorems involving one spin-3/2 and one
spin-1/2 spherical harmonic.

\paragraph*{$\boldsymbol{|\Delta\ell|=1}$}
\label{sec:3h1h-2}

We consider the case $s'=3/2$, $s=1/2$ first.  In this case
$S(j,\ell',\ell,3/2,1/2)$ can be non-vanishing only if 
$\ell'=\ell+1$ and $j=\ell\pm1/2$, or if $\ell'=\ell-1$ and
$j=\ell\pm1/2$.  With the spin matrix elements in (\ref{eq:3h1h-1}),
from (\ref{eq:CGmaster}) we obtain
\begin{subequations}
  \label{eq:3h1h-7}
\begin{equation}
  \label{eq:3h1h-7a}
  \begin{aligned}
    S(j,\ell_1,\ell,3/2,1/2) &= \frac{1}{2} \sqrt{\frac{3}{2}}
    C^{\frac{3}{2}\frac{1}{2}1}_{\ell_1\ell j} \langle\ell_1\ell'_z |
    \veps{k}(\dl_z)\verr{k} | \ell,\ell_z\rangle
    \veps{h}(\dl_z)^* \vchiU{A}{h}(s'_z)^* \vchi{A}(s_z)\\
    &+ \frac{1}{4} \sqrt{\frac{3}{2}}
    C^{\frac{3}{2}\frac{1}{2}2}_{\ell_1\ell j} \langle\ell_1\ell'_z |
    \veps{k_1k_2}(\dl_z)\verr{k_1} L^{k_2} | \ell,\ell_z\rangle
    \veps{h_1h_2}(\dl_z)^* \vchiU{A}{h_1}(s'_z)^*
    \sigma^{h_2}_{AB}\vchi{B}(s_z)~,
  \end{aligned}
\end{equation}
with the coefficients
\begin{equation}
  \label{eq:3h1h-7b}  
\begin{aligned}
  C^{\frac{3}{2}\frac{1}{2}1}_{\ell_1\ell j} &=-\sqrt{2} \ds \dl
  \sqrt{\frac{2j+1}{2\lave+1}} \sqrt{\frac{2j-\lave+\ds\dl
      +\frac{1}{2}} {2\lave-\ds\dl +1}}
  \sqrt{1-2\ds\dl (j-\lave)}~,\\
C^{\frac{3}{2}\frac{1}{2}2}_{\ell_1\ell j} &=-\frac{8}{\sqrt{3}} \sqrt{
  \frac{\frac{3}{2}+\ds\dl(j-\lave)} {2\lave-\ds\dl + 1}}\,
\frac{1+2\ds\dl (j-\lave)} {\sqrt{j+\frac{1}{2}+\ds \dl}}~,
  \end{aligned}
\end{equation}
\end{subequations}
where in this case $\ds=1$. 
On the first line of (\ref{eq:3h1h-7a}) we can replace
$\veps{k}(\dl_z) \veps{h}(\dl_z)^*$ with $\delta^{kh}$, and on the
second line,
\begin{equation}
  \label{eq:3h1h-8}
  \begin{split}
    \langle\ell'\ell'_z | \veps{k_1k_2}(\dl_z)\verr{k_1} L^{k_2} |
    \ell,\ell_z\rangle \veps{h_1h_2}(\dl_z)^* \vchiU{A}{h_1}(s'_z)^*
    \sigma^{h_2}_{AB}\vchi{B}(s_z) = \hspace{35ex}\\
    =\frac{1}{2}
    \langle\ell'\ell'_z | \verr{\{k_1} L^{k_2\}_0} |
    \ell,\ell_z\rangle  \vchiU{A}{h_1}(s'_z)^*
    \sigma^{h_2}_{AB}\vchi{B}(s_z)~.
  \end{split}
\end{equation}
Similarly, from (\ref{eq:6j-2}) we get
\begin{equation}
  \label{eq:3h1h-9}
  \begin{split}
  S(j,\ell_1,\ell,3/2,1/2) = -\frac{1}{2}\sqrt{\frac{3}{2}}\dl
  \sqrt{\frac{2\lave+1}{2\ell_1+1}} C^{\frac{3}{2}\frac{1}{2}1}_{\ell_1\ell j}
  \CG{\ell}{\ell_z}{1}{\dl_z}{\ell_1}{\ell'_z}
  \CG{3/2}{s'_z}{1}{\dl_z}{1/2}{s_z}
  \hspace{12ex}\\ 
  - \frac{\sqrt{30}}{32} \dl  \sqrt{\frac{2\lave+1}{2\ell_1+1}}
  \sqrt{(2\lave-1)(2\lave+3)} C^{\frac{3}{2}\frac{1}{2}2}_{\ell_1\ell j}
  \CG{\ell}{\ell_z}{2}{\dl_z}{\ell_1}{\ell'_z}
  \CG{3/2}{s'_z}{2}{\dl_z}{1/2}{s_z}~,
  \end{split}  
\end{equation}
with the coefficients (\ref{eq:3h1h-7b}).

In the case $s'=1/2$, $s=3/2$, $S(j,\ell',\ell,1/2,3/2)$ can be
non-vanishing only if $\ell'=\ell+1$ and $j=\ell+3/2$, $\ell+1/2$, 
or if $\ell'=\ell-1$ and $j=\ell-1/2$, $\ell-3/2$.  The treatment of
this case is completely analogous to the previous one. In terms of
matrix elements, from (\ref{eq:CGmaster}) we obtain 
\begin{equation}
  \label{eq:3h1h-a}
  \begin{aligned}
  S(j,\ell_1,\ell,1/2,3/2) &= \frac{1}{2} \sqrt{\frac{3}{2}}
  C^{\frac{1}{2}\frac{3}{2}1}_{\ell_1\ell j}  
  \langle\ell_1\ell'_z | \veps{k}(\dl_z)\verr{k} | \ell,\ell_z\rangle 
  \veps{h}(\dl_z)^* \vchi{A}(s'_z)^* \vchiU{A}{h}(s_z)\\
  &+ \frac{1}{4} \sqrt{\frac{3}{2}}
  C^{\frac{1}{2}\frac{3}{2}2}_{\ell_1\ell j} 
  \langle\ell_1\ell'_z | \veps{k_1k_2}(\dl_z) \verr{k_1} L^{k_2} |
  \ell,\ell_z\rangle \veps{h_1h_2}(\dl_z)^* \vchi{A}(s'_z)^*
  \sigma^{h_1}_{AB}\vchiU{B}{h_2}(s_z)~, 
  \end{aligned}
\end{equation}
with the coefficients $C^{\frac{1}{2}\frac{3}{2}\Delta}_{\ell_1\ell
  j}$ given by (\ref{eq:3h1h-7b}) with $\ds=-1$.  From (\ref{eq:6j-2})
we get, analogously,
\begin{equation}
  \label{eq:3h1h-b}
  \begin{split}
  S(j,\ell_1,\ell,1/2,3/2) = \frac{\sqrt{3}}{4}\dl
  \sqrt{\frac{2\lave+1}{2\ell_1+1}} C^{\frac{1}{2}\frac{3}{2}1}_{\ell_1\ell j}
  \CG{\ell}{\ell_z}{1}{\dl_z}{\ell_1}{\ell'_z}
  \CG{1/2}{s'_z}{1}{\dl_z}{3/2}{s_z}
  \hspace{12ex}\\ 
  + \frac{\sqrt{15}}{32} \dl  \sqrt{\frac{2\lave+1}{2\ell_1+1}}
  \sqrt{(2\lave-1)(2\lave+3)} C^{\frac{1}{2}\frac{3}{2}2}_{\ell_1\ell j}
  \CG{\ell}{\ell_z}{2}{\dl_z}{\ell_1}{\ell'_z}
  \CG{1/2}{s'_z}{2}{\dl_z}{3/2}{s_z}~.
  \end{split}  
\end{equation}
The equalities (\ref{eq:3h1h-7}) and (\ref{eq:3h1h-a}) will
be used for the derivation of an addition theorem involving one
spin-3/2 and one spin-1/2 spherical harmonic, with orbital angular
momentum differing by one unit.

\paragraph*{$\boldsymbol{|\Delta\ell|=2}$}
\label{sec:3h1h-3}

As in the previous paragraph, we consider the case $s'=3/2$, $s=1/2$
first.  In this case $S(j,\ell',\ell,3/2,1/2)$ can be non-vanishing
only if $\ell'=\ell+2$ and $j=\ell+1/2$, or if $\ell'=\ell-2$ and
$j=\ell-1/2$.  Using the spin matrix elements (\ref{eq:3h1h-1}) and
the range of $j$, (\ref{eq:CGmaster}) reduces to
\begin{subequations}
  \label{eq:3h1h-c}
\begin{align}
  \label{eq:3h1h-ca}
  S(j,\ell_2,\ell,3/2,1/2) &= \frac{1}{4} \sqrt{\frac{3}{2}}
  C^{\frac{3}{2}\frac{1}{2}2}_{\ell_2\ell j} 
  \langle\ell_2\ell'_z | \veps{k_1k_2}(\dl_z) \verr{k_1} \verr{k_2} | 
  \ell,\ell_z\rangle \veps{h_1h_2}(\dl_z)^*
  \vchiU{A}{h_1}(s'_z)^* \sigma^{h_2}_{AB} \vchi{B}(s_z)~,  \\
  \label{eq:3h1h-cb}
  C^{\frac{3}{2}\frac{1}{2}2}_{\ell_2\ell j} &= \frac{2}{\sqrt{3}} \ds
  \dl \sqrt{2j+1} \sqrt{\frac{2\lave+1}{\lave(\lave+1)}}~,
\end{align}
\end{subequations}
where $\ds=1$.  The matrix elements in (\ref{eq:3h1h-ca}) can be
written without standard tensors as,
\begin{equation}
  \label{eq:3h1h-cx}\hspace*{-7pt}
  \langle\ell_2\ell'_z | \veps{k_1k_2}(\dl_z) \verr{k_1} \verr{k_2} | 
  \ell,\ell_z\rangle \veps{h_1h_2}(\dl_z)^*
  \vchiU{A}{h_1}(s'_z)^* \sigma^{h_2}_{AB} \vchi{B}(s_z)  = 
  \langle\ell_2\ell'_z |  \verr{k_1} \verr{k_2} | 
  \ell,\ell_z\rangle 
  \vchiU{A}{k_1}(s'_z)^* \sigma^{k_2}_{AB} \vchi{B}(s_z).
\end{equation}
Similarly, in this case (\ref{eq:6j-2}) takes the
simplified form 
%
\begin{equation}
  \label{eq:3h1h-d}
  \begin{aligned}
  S(j,\ell_2,\ell,3/2,1/2) &= -\frac{\sqrt{30}}{8} \sqrt{2\ell+1}
  \sqrt{\frac{\lave (\lave+1)}{(2\lave-1)(2\lave+1)(2\lave+3)}}
  C^{\frac{3}{2}\frac{1}{2}2}_{\ell_2\ell j} \\
  &\quad\times \CG{\ell}{\ell_z}{2}{\dl_z}{\ell_2}{\ell'_z}
  \CG{3/2}{s'_z}{2}{\dl_z}{1/2}{s_z}~.
  \end{aligned}
\end{equation}

In the case $s'=1/2$, $s=3/2$, $S(j,\ell',\ell,1/2,3/2)$ can be
non-vanishing only if $\ell'=\ell+2$ and $j=\ell'-1/2$, or if 
$\ell'=\ell-2$ and $j=\ell'+1/2$.  In this case (\ref{eq:CGmaster})
reduces to
\begin{equation}
\begin{aligned}
  \label{eq:3h1h-e}
  S(j,\ell_2,\ell,1/2,3/2) &= \frac{1}{4} \sqrt{\frac{3}{2}}
  C^{\frac{1}{2}\frac{3}{2}2}_{\ell_2\ell j} 
  \langle\ell_2\ell'_z | \veps{k_1k_2}(\dl_z) \verr{k_1} \verr{k_2} | 
  \ell,\ell_z\rangle \veps{h_1h_2}(\dl_z)^*
  \vchi{A}(s'_z)^* \sigma^{h_2}_{AB} \vchiU{B}{h_1}(s_z)~,  \\
\end{aligned}
\end{equation}
with $C^{\frac{1}{2}\frac{3}{2}2}_{\ell_2\ell j}$ given by the same
expression (\ref{eq:3h1h-cb}) as
$C^{\frac{3}{2}\frac{1}{2}2}_{\ell_2\ell j}$, with $\ds=-1$.  The
equality (\ref{eq:6j-2}) takes the simplified form
%
\begin{equation}
\begin{aligned}
  \label{eq:3h1h-f}
  S(j,\ell_2,\ell,1/2,3/2) &= \frac{\sqrt{15}}{8} \sqrt{2\ell+1} 
  \sqrt{\frac{\lave(\lave+1)}{(2\lave-1)
    (2\lave+1)(2\lave+3)}}C^{\frac{1}{2}\frac{3}{2}2}_{\ell_2\ell j} \\
  &\times \CG{\ell}{\ell_z}{2}{\dl_z}{\ell_2}{\ell'_z}
  \CG{1/2}{s'_z}{2}{\dl_z}{3/2}{s_z}~,
\end{aligned}
\end{equation}

\section{Angular momentum projection operators}
\label{sec:proj}

We define the angular momentum projection operators as
\begin{equation}
  \label{eq:proj}
\P_\ell = \sum_{m=-\ell}^\ell |\ell,m\rangle\langle\ell,m|~,
\qquad
0\leq\ell<\infty~.  
\end{equation}
It is immediate that $\P_\ell$ are a sequence of orthogonal
projectors, which commute with $\vec{L}$ and resolve the identity, 
\begin{equation}
  \label{eq:id}
  \P_{\ell'}\P_\ell = \P_\ell\delta_{\ell'\ell}~,
\qquad
\sum_{\ell=0}^\infty \P_\ell = 1~,
\qquad
[\vec{L},\P_\ell]=0~.
\end{equation}
The matrix elements of $\P_\ell$ in configuration space follow from
their definition (\ref{eq:proj}) and the addition theorem for
spherical harmonics
\begin{equation}
  \label{eq:proj3}
  \P_\ell|\verr{}\rangle = \sum_{m=-\ell}^\ell Y_{\ell
    m}(\verr{})^*|\ell,m\rangle~, 
  \qquad
  \langle\verrp{}|\P_\ell|\verr{}\rangle = \frac{2\ell+1}{4\pi}
  P_\ell(\rdotrp)~. 
\end{equation}
From (\ref{eq:proj})--(\ref{eq:proj3}) the matrix elements of
$\P_\ell$ with other orbital operators can be computed.  Those matrix
elements, together with the results on products of CG coefficients of
the previous section, are the basic building blocks needed in our
derivation of addition theorems for spin spherical harmonics. As
will be discussed in II, however, the results for $\P_\ell$
matrix elements constitute by themselves addition theorems for
spherical harmonics. 

\subsection{Matrix elements with tensor powers of $\boldsymbol{\vec{L}}$}
\label{sec:Lpower}

The matrix elements of $\P_\ell$ with tensor powers of $\vec{L}$ will
be used in the derivation of addition theorems for spin spherical
harmonics.  For the irreducible components, from which the full tensor
matrix elements can be reconstructed, we have
\begin{equation}
  \label{eq:recL}
  \begin{aligned}
  \langle\verrp{}|L^{\{h_1}\ldots L^{h_p\}_0} \P_\ell|\verr{}\rangle &=
  \frac{2\ell+1}{4\pi}
  i^p \left(\verr{}\wedge\nabla\right)^{\{h_p}
  \ldots \left(\verr{}\wedge\nabla \right)^{h_1\}_0}
  P_\ell(x) \\
  &= i \left(\verr{}\wedge\nabla\right)^{\{h_p}
    \langle\verrp{}|L^{h_1}\ldots L^{h_{p-1}\}_0}
    \P_\ell|\verr{}\rangle \\
    &=\frac{i}{(p-1)!} \left(\verr{}\wedge\nabla\right)^{\{h_p}
    \langle\verrp{}|L^{\{h_1}\ldots L^{h_{p-1}\}_0\}_0}
    \P_\ell|\verr{}\rangle~.
  \end{aligned}
\end{equation}
Thus, we can obtain these irreducible matrix elements recursively,
with the result
\begin{subequations}
  \label{eq:genL}  
\begin{equation}
  \label{eq:genLa}  
  \langle\verrp{}|L^{\{h_1}\ldots L^{h_p\}_0} \P_\ell|\verr{}\rangle 
  = \frac{2\ell+1}{4\pi} i^p \sum_{q=0}^{[p/2]}
  C_{p,q} Z^{\{h_1\ldots h_q;h_{q+1}\ldots h_{2q}}
  v^{h_{2q+1}}\ldots v^{h_p\}_0} P_\ell^{(p-q)}(x),
\end{equation}
with $P_\ell^{(k)}$ the $k^\mathrm{th}$ derivative of $P_\ell$ and,
\begin{equation}
  \label{eq:genLb}  
\begin{gathered}
  x=\rdotrp~,\quad \vec{v} = \verr{}\wedge\verrp{}~,\quad
  C_{p,q} = ( 2q-1)!! \binom{p}{2q}~,\quad
  C_{p,0} = 1~,\\
  Z^{i_1\ldots i_{n_1};j_1\ldots j_{n_2}} = \verr{i_1} \ldots
  \verr{i_{n_1}} \verrp{j_1} \ldots
  \verrp{j_{n_2}} ~.
\end{gathered}
\end{equation}
For notational simplicity in what follows we adopt the following
useful conventions.  In $Z^{i_1\ldots i_{n_1};j_1\ldots j_{n_2}}$ one
or both sets of indices may be empty,
\begin{equation}
  \label{eq:genLbaux}
  Z^{\mbox{ };j_1\ldots j_{n_2}} = \verrp{j_1} \ldots
  \verrp{j_{n_2}}~,
  \quad
  Z^{i_1\ldots i_{n_2};\mbox{ }} = \verr{i_1} \ldots
  \verr{i_{n_1}}~,
  \quad
  Z^{\mbox{ };\mbox{ }} = 1~.  
\end{equation}
In the term $q=0$ in (\ref{eq:genLa}) the index sets are empty,
$Z^{h_1\ldots h_0;h_{1}\ldots h_{0}} \equiv Z^{\mbox{ };\mbox{ }}$.
Similarly, in the term $q=[p/2]$, when $p$ is even, we set
$v^{h_{2q+1}}\ldots v^{h_p} = v^{h_{p+1}}\ldots v^{h_p} \equiv 1$.
\end{subequations}

The cases $n=1$, 2, 3 will be needed for the derivation of addition
theorems for spherical harmonics of spin 1/2, 1 and 3/2 so we quote
them here explicitly,
\begin{subequations}
\label{eq:Ls}
\begin{align}
  \label{eq:L}
  \langle\verrp{}|\vec{L}\P_\ell|\verr{}\rangle &= i
  \frac{2\ell+1}{4\pi} \vec{v} P'_\ell(\rdotrp)~, \\
  \label{eq:LL}
  \langle\verrp{}|L^{\{i} L^{j\}_0} \P_\ell|\verr{}\rangle
  &= -\frac{2\ell+1}{4\pi} \left( v^{\{i} v^{j\}_0} P''_\ell(\rdotrp)
    + \verr{\{i} \verrp{j\}_0} P'_\ell(\rdotrp) 
  \right)~,\\
  \label{eq:LLL}
  \langle\verrp{}|L^{\{i} L^{j} L^{k\}_0} \P_\ell|\verr{}\rangle
  &= -i \frac{2\ell+1}{4\pi} \left( v^{\{i} v^{j} v^{k\}_0}
    P'''_\ell(\rdotrp) + 3 \verrp{\{i} \verr{j} v^{k\}_0}
    P''_\ell(\rdotrp) \right)~. 
\end{align}
\end{subequations}

\subsection{Matrix elements with tensor powers of $\boldsymbol{\verr{}}$}
\label{sec:rpower}

The matrix element $\langle
\verpp{}|\P_{\ell'}\verr{i_1}\ldots\verr{i_n}\P_\ell | \verp{}\rangle$
is an irreducible tensor if $\ell'=\ell\pm n$, otherwise reducible.
The irreducible matrix elements will be used in the derivation of
addition theorems for spin spherical harmonics.  They satisfy the
recursion relation,
\begin{multline}
  \label{eq:rmat1}  
  \langle\verpp{} |
  \P_{\ell_{n+1}}\verr{i_1}\ldots\verr{i_{n+1}}\P_\ell |
  \verp{}\rangle =\\
  -\frac{1}{2\ell_n+1} \left[ (\ell_{n+1}-\ell_n)
    \nabla^{i_{n+1}}_{p'} - \frac{1}{2} (\ell_{n+1} + \ell_{n}
    +1) \verpp{k} \right]
  \langle\verpp{}|\P_{\ell_{n}}\verr{i_1}\ldots\verr{i_{n}}\P_\ell |
  \verp{}\rangle~,
\end{multline}
with
\begin{equation}
  \label{eq:rmat2}
    \langle\verpp{} |
  \P_{\ell_{1}}\verr{}\P_\ell |\verp{}\rangle = 
  \frac{\ell_1 - \ell}{4\pi} \left( \verpp{} P'_{\ell_1}(x) - \verp{}
    P'_\ell(x) \right)~. 
\end{equation}
In these equations we used the notation $\ell_k = \ell\pm k$, see
appendix \ref{sec:notatio}.  Eqs.\ (\ref{eq:rmat1}), (\ref{eq:rmat2})
are established in appendix \ref{sec:appa}.
The recursion relation (\ref{eq:rmat1}) can be solved with the initial
condition (\ref{eq:rmat2}) to give
\begin{equation}
    \label{eq:rmat3}
\begin{gathered}
  \langle\verpp{}|\P_{\ell_{n}}\verr{i_1}\ldots\verr{i_{n}}\P_\ell |
  \verp{}\rangle = A_n
  \sum^n_{\substack{k_1,k_2=0\\k_1+k_2=n}} \frac{(-1)^{k_2}}{k_1!k_2!}
  Z^{\{i_1\ldots i_{k_2};i_{k_2+1}\ldots i_n\}_0} P_{\ell_{k_1}}^{(n)}(x)~,\\
  A_n = \frac{(\ell_1-\ell)^n}{4\pi} \frac{2\ell+1}{\prod_{k=0}^{n-1}
    (2\ell_k+1)}
  = \frac{1}{4\pi} \left(\frac{\ell_n-\ell}{n}\right)^n (2\ell+1) 
  \frac{(\ell_n+\ell-n-\frac{\ell_n-\ell}{n})!!}
  {(\ell_n+\ell+n-\frac{\ell_n-\ell}{n})!!}~,
\end{gathered}
\end{equation}
with $Z^{i_1\ldots i_{n_1};j_{1}\ldots j_{n_2}}$ defined in
(\ref{eq:genLb}), (\ref{eq:genLbaux}).  As before, we adopt the
convention that in the term with $k_1=n$, $k_2=0$ in (\ref{eq:rmat3})
the first index set in $Z$ is empty, $Z^{i_1\ldots i_0; i_1\ldots i_n}
\equiv Z^{\mbox{ }; i_1\ldots i_n}$ and, in the term with $k_1=0$,
$k_2=n$, the second index set is empty $Z^{i_1\ldots i_n;
  i_{n+1}\ldots i_n} \equiv Z^{i_1\ldots i_n; \mbox{ }}$.

The particular cases $n=1$, 2, 3 will be needed for the derivation of
addition theorems for spherical harmonics of spin 1/2, 1 and 3/2, so
we give them here explicitly,
\begin{subequations}
  \label{eq:rmat4}
  \begin{align}
    \label{eq:rmat4a}
  \langle\verpp{}|\P_{\ell_{1}}\verr{i}\P_\ell |\verp{}\rangle &= 
  \frac{\ell_1-\ell}{4\pi} \left(-\verp{i} P'_\ell(x) + \verpp{i}
    P'_{\ell_1}(x) \right),\\
    \label{eq:rmat4b}
  \langle\verpp{}|\P_{\ell_{2}}\verr{i_1}\verr{i_{2}}\P_\ell |
  \verp{}\rangle &=
  \frac{1}{4\pi} \frac{1}{2\ell_1+1} \left(
    \frac{1}{2} \verp{\{i_1} \verp{i_2\}_0} P''_\ell(x) -
    \verp{\{i_1} \verpp{i_2\}_0} P''_{\ell_1}(x)
    +\frac{1}{2} \verpp{\{i_1} \verpp{i_2\}_0} P''_{\ell_2}(x)
  \right),\\
  \nonumber
  \langle\verpp{}|\P_{\ell_{3}}\verr{i_1}\verr{i_{2}}
  \verr{i_{3}}\P_\ell | \verp{}\rangle &=
  \frac{\ell_1-\ell}{4\pi} \frac{1}{(2\ell_2+1)(2\ell_1+1)} \left(
    -\frac{1}{6} \verp{\{i_1} \verp{i_2} \verp{i_3\}_0} P'''_\ell(x)
    +\frac{1}{2} \verp{\{i_1} \verp{i_2} \verpp{i_3\}_0}
    P'''_{\ell_1}(x)\right.\\
    \label{eq:rmat4c}
    &\quad \left.
    -\frac{1}{2} \verp{\{i_1} \verpp{i_2} \verpp{i_3\}_0}
    P'''_{\ell_2}(x) 
    +\frac{1}{6} \verpp{\{i_1} \verpp{i_2} \verpp{i_3\}_0} P'''_{\ell_3}(x)
  \right).
  \end{align}
\end{subequations}

\subsection{Mixed matrix elements}
\label{sec:mix}

We consider now matrix elements of $\P_\ell$ with tensor products of
both $\vec{L}$ and $\verr{}$.  The matrix elements of the irreducible
tensor operator $\verr{\{i_1}\ldots\verr{i_n} L^{h_1}\ldots
L^{h_p\}_0}$ are of the form
\begin{equation}
  \label{eq:mix1}
  \langle\verpp{} | \P_{\ell_{n}}\verr{\{i_1}\ldots\verr{i_{n}}
  L^{h_1}\ldots L^{h_t\}_0}\P_\ell |\verp{}\rangle = i^t
  (\verp{}\wedge\nabla_p)^{\{h_t} \ldots (\verp{}\wedge\nabla_p)^{h_1} 
  \langle\verpp{} | \P_{\ell_{n}}\verr{i_1}\ldots\verr{i_{n}\}_0}
  \P_\ell |\verp{}\rangle ~.
\end{equation}
From this equation and (\ref{eq:rmat3}), for $t=1$ we obtain
\begin{equation}
  \label{eq:mix2}
  \langle\verpp{}|\P_{\ell_{n}}\verr{\{i_1}\ldots\verr{i_{n}}L^{h\}_0}
  \P_\ell | \verp{}\rangle = i A_n
  \sum^n_{\substack{k_1,k_2=0\\k_1+k_2=n}} (-1)^{k_2} \binom{n}{k_1}
  Z^{\{i_1\ldots i_{k_2};i_{k_2+1}\ldots i_n}
  v^{h\}_0}P_{\ell_{k_1}}^{(n+1)}(x)~.
\end{equation}
For $s>1$, equation (\ref{eq:mix1}) leads to the recursion relation 
\begin{equation}
  \label{eq:mix3}
  \begin{split}
  \langle\verpp{} | \P_{\ell_{n}}\verr{\{i_1}\ldots\verr{i_{n}}
  L^{h_1}\ldots L^{h_t\}_0}\P_\ell |\verp{}\rangle = \hspace{60ex}\\
  =\frac{i}{(n+t-1)!}
  (\verp{}\wedge\nabla_p)^{\{h_t}
  \langle\verpp{} | \P_{\ell_{n}}\verr{\{i_1}\ldots\verr{i_{n}}
  L^{h_1}\ldots L^{h_{t-1}\}_0\}_0} \P_\ell |\verp{}\rangle ~,  
  \end{split}
\end{equation}
which can be solved with the initial condition (\ref{eq:mix2}) to
yield 
\begin{equation}
  \label{eq:mix4}
  \begin{split}
  \langle\verpp{} | \P_{\ell_{n}}\verr{\{i_1}\ldots\verr{i_{n}}
  L^{h_1}\ldots L^{h_t\}_0}\P_\ell |\verp{}\rangle = \hspace{63ex}\\
  =i^t A_n 
  \sum^n_{\substack{k_1,k_2=0\\k_1+k_2=n}} (-1)^{k_2} \binom{n}{k_1}
  \sum_{q=0}^{[t/2]} C_{t,q} 
  Z^{\{i_1\ldots i_{k_2}h_1\ldots h_q;i_{k_2+1}\ldots i_n
    h_{q+1}\ldots h_{2q}} v^{h_{2q+1}}\ldots v^{h_t\}_0}
  P_{\ell_{k_1}}^{(n+t-q)}(x)~,
\end{split}
\end{equation}
with $[\mbox{ }]$ in the inner summation denoting integer part, $A_n$
as defined in (\ref{eq:rmat3}), and $C_{s,q}$, $x$, $\vec{v}$, and
$Z^{i_1\ldots i_{n_1};j_{1}\ldots j_{n_2}}$ as defined in
(\ref{eq:genLb}), with $\verp{}$ instead of $\verr{}$.  As before, for
notational simplicity, we have not separated from the sum the terms
with $(k_1,k_2)=(n,0)$, $(0,n)$, those with $q=0$ and, for even $s$,
those with $q=[s/2]$.  In those cases we apply the same conventions as
explained after eqs.\ (\ref{eq:genL}) and (\ref{eq:rmat3}).

Particular cases of importance for the derivation of addition theorems
for spherical harmonics of spin 1/2, 1 and 3/2, are
\begin{subequations}
  \label{eq:mix5}
\begin{align}
  \langle\verpp{} | \P_{\ell_{1}}\verr{\{i}L^{k\}}\P_\ell |\verp{}\rangle
  &= i \frac{\ell_1-\ell}{4\pi} \left(\verpp{\{i}v^{k\}}
    P_{\ell_1}''(x) - \verp{\{i}v^{k\}} P_{\ell}''(x)
  \right),\label{eq:mix5a}\\
  \langle\verpp{}|\P_{\ell_{2}}\verr{\{i}\verr{j}L^{k\}_0}
  \P_\ell | \verp{}\rangle &= \frac{i}{4\pi} \frac{1}{2\ell_1+1}
  \left(
    \verpp{\{i}\verpp{j}v^{k\}_0} P'''_{\ell_2}(x) -
    2\verpp{\{i}\verp{j}v^{k\}_0} P'''_{\ell_1}(x) +
    \verp{\{i}\verp{j}v^{k\}_0} P'''_{\ell}(x)
  \right),\label{eq:mix5b}\\
  \langle\verpp{} | \P_{\ell_{1}}\verr{\{i}L^{j_1}L^{j_2\}_0}\P_\ell |\verp{}\rangle
  &= - \frac{\ell_1-\ell}{4\pi} \left(\verpp{\{i}v^{j_1}v^{j_2\}_0}
    P_{\ell_1}'''(x) - \verp{\{i}v^{j_1}v^{j_2\}_0} P_{\ell}'''(x) 
  + \verpp{\{i}\verpp{j_1}\verp{j_2\}_0} P''_{\ell_1}(x) 
  \right.  \nonumber\\  
  &\qquad\left. - 
    \verpp{\{i}\verp{j_1}\verp{j_2\}_0} P''_{\ell}(x) \right).\label{eq:mix5c}
\end{align}
\end{subequations}

\section{Final remarks}
\label{sec:finrem}

We have presented in the foregoing sections preliminary results needed
for the systematic derivation of addition theorems for spin spherical
harmonics in II.  In sect.\ \ref{sec:fac} we obtained the
factorization of orbital and spin degrees of freedom in products of CG
coefficients of the form (\ref{eq:additionintro}) in general form, and
discussed the particular cases with $0\leq s',s\leq 3/2$.  In those
cases the coefficients $C^{s's\Delta}_{\ell'\ell j}$, given in full
generality in (\ref{eq:CGmaster}), were reduced to much smaller forms,
and the tensor and spinor structures of each term in the expansion
given explicitly.

In sect.\ \ref{sec:proj} the matrix elements of the angular-momentum
projector operator with the irreducible components of arbitrary tensor
products of $\verr{}$ and $\vec{L}$ are given in general form, for
all values of their parameters.  Those matrix elements will be used in
II to obtain general expressions for bilocal spherical harmonics, and
to derive addition theorems for spin spherical harmonics.  Their
applicability is even wider, however, when they are appropriately
combined to obtain matrix elements of reducible tensor operators.
Some examples of those applications will also be considered in II.

As a side remark we point out that an unexpected byproduct of the
results of sect.\ \ref{sec:fac} is an improvement in computational
efficiency, at least in the specific case of infinite-precision
computation \cite{wolf} in which the results are given as a rational
number times the square root of a rational number.  Consider, for
example, the computation of both sides of (\ref{eq:3h-3}), at fixed
$\ell$ and $j$, for all possible values of $-\ell\leq
\ell'_z,\ell_z\leq \ell$ and $-3/2\leq s'_z, s_z\leq 3/2$.  If the
l.h.s.\ of (\ref{eq:3h-3}) is computed as written on the second line
of (\ref{eq:additionintro}), we find that the ratio of CPU times
$\tau_\mathrm{r.h.s.}/\tau_\mathrm{l.h.s.}$ begins at $\sim 2$ at
$\ell=1$, monotonically decreasing with $\ell$ to reach 1 at $\ell\sim
20$, $\sim 0.5$ at $\ell\sim100$, and $\sim 0.25$ at $\ell\sim200$.
We remark that those numbers are subject to statistical fluctuations.

\appendix

\renewcommand{\theequation}{\thesection.\arabic{equation}}
\setcounter{equation}{0}

\section{Notation and conventions}
\label{sec:notatio}

Throughout the paper we set $\hbar=1$.  We denote tensor product
states of orbital and spin angular momentum by $|\ell, \ell_z; s, s_z
\rangle = |\ell, \ell_z\rangle \otimes | s, s_z \rangle $, and states
coupled to total angular momentum $j$, $j_z$ by $|\ell, s, j,
j_z\rangle$. We follow the notation of \cite{ham89} for CG
coefficients, 
\[
\CG{\ell}{\ell_z}{s}{s_z}{j}{j_z} \equiv \langle\ell, \ell_z; s, s_z |
\ell, s, j, j_z\rangle~.
\]  
(Different notations are used, e.g., in \cite{edm96,var88,pdg,gal90}.)
We adopt the usual convention that $\CG{\ell}{\ell_z}{s}{s_z}{j}{j_z}
= 0$ if $\ell_z+s_z \neq j_z$, or $j<|\ell-s|$, or $j>\ell+s$.  For
Wigner $6j$-symbols \cite{edm96,bie85a,bie85b,var88} we use the
definition and notation of \cite{edm96,var88}.  The most important
property of $6j$-symbols for our purposes is the relation \cite{edm96}
\begin{equation}
  \label{eq:6j-1}
  \begin{aligned}
\lefteqn{\CG{j_1}{m_1}{j_2}{m_2}{j'}{m_1+m_2}
  \CG{j'}{m_1+m_2}{j_3}{m_3}{j}{m} = 
  \sum_{j''=|j_2-j_3|}^{j_2+j_3} \sqrt{(2j'+1)(2j''+1)}
  (-1)^{j_1+j_2+j_3+j} }\hspace{27ex}
  \\
  &\times \SJ{j_1}{j_2}{j'}{j_3}{j}{j''}
  \CG{j_1}{m_1}{j''}{m_2+m_3}{j}{m}
  \CG{j_2}{m_2}{j_3}{m_3}{j''}{m_2+m_3}.
  \end{aligned}
\end{equation}
This relation is satisfied by the implementation of CG coefficients
and $6j$-symbols in the software system \cite{wolf}.  Our notation for
the reduced matrix elements entering the WE theorem is explained in
appendix \ref{sec:tensor}.  For spherical harmonics $Y_{\ell m}$ and
Legendre polynomials $P_\ell$ we use the standard definitions
\cite{var88,edm96,gal90}.  Furthermore, we adopt the convention, usual
in numerical computations \cite{wolf}, $Y_{\ell m}(\verr{})\equiv 0$
if $|m|>\ell$. When states with different angular momenta or spins are
considered, we use the notations,
\begin{equation}
  \label{eq:notatio1}
  \lave = \frac{\ell'+\ell}{2}~, 
\quad
  \dl = \ell'-\ell~,
\quad
  \dl_z= \ell'_z-\ell_z~,
\end{equation}
and similarly $\save$, $\ds$, $\ds_z$.  We also find useful the
notation $\ell_k = \ell\pm k$ for orbital angular-momentum quantum
numbers, with the convention that when several $\ell_{k_1}$, \ldots,
$\ell_{k_p}$ appear in the same equation, then either the upper or the
lower sign is chosen for all of them simultaneously.  Thus,
$\ell_{n+1}-\ell_{n} = \ell_{p+1}-\ell_{p} =\pm 1$ for any $n$ and
$p$.

We denote tensor indices by lower-case latin superindices,
$A^{i_1\ldots i_n}$, and spinor indices by upper-case latin
subindices, $\chi_{A}$.  Since we consider only tensors and spinors of
the su(2) algebra (as opposed to the sl(2) algebra), there is no need
to raise spinor indices or lower tensor ones.  We denote normal
spinors and tensors by a caret, e.g., $\verr{}=\vec{r}/r$.
Our choice of orthonormal bases for spinors and tensors is
explained in appendix~\ref{sec:standard}.

Given a numeric or operator tensor $A^{i_1\ldots i_n}$, we denote its
associated totally symmetrized and antisymmetrized tensors as,
\begin{equation}
  \label{eq:notatio2}
  A^{\{i_1\ldots i_n\}} = \sum_{\sigma} A^{i_{\sigma_1}\ldots
    i_{\sigma_n}}~,
  \qquad
  A^{[i_1\ldots i_n]} = \sum_{\sigma} \mathrm{sgn}(\sigma)A^{i_{\sigma_1}\ldots
    i_{\sigma_n}}~,
\end{equation}
where the sums extend over all permutations $\sigma$ of
$i_1,\ldots,i_n$.  The same notations apply to tensor products, e.g.,
$\verr{\{i}L^{j\}} = \verr{i}L^j + \verr{j} L^i$.  We denote the
traceless part of $A^{i_1\ldots i_n}$ by $A^{(i_1\ldots i_n)_0}$.  The
cases of importance in this paper are rank-2 and 3 tensors,
\begin{equation}
  \begin{aligned}
  A^{(ij)_0} &= A^{ij} - \frac{1}{3} A^{kk} \delta^{ij}~,\\
  A^{(pqr)_0} &= A^{pqr} -\frac{2}{5}(A^{pjj}\delta^{qr} + A^{iqi}\delta^{pr} +
  A^{iir}\delta^{pq}) \\
  &\quad + \frac{1}{10} (A^{qjj}\delta^{pr} +
  A^{rjj}\delta^{pq} + A^{ipi}\delta^{qr} + A^{iri}\delta^{pq} +
  A^{iip}\delta^{qr} + A^{iiq}\delta^{pr})~.
  \end{aligned}
\end{equation}
The same notation applies to tensor products, such as $\verr{(i}
\verr{j)_0} = \verr{i}\verr{j}-1/3 \delta^{ij}$.  We denote by
$A^{\{i_1\ldots i_n\}_0}$ the traceless part of $A^{\{i_1\ldots
  i_n\}}$.  Thus, the irreducible component of $A^{i_1\ldots i_n}$ is
$1/n! A^{\{i_1\ldots i_n\}_0}$~.
A more systematic approach
to irreducible tensors is given in the following section.

\setcounter{equation}{0}
\section{Standard tensor and spinor bases}
\label{sec:standard}

In this appendix we introduce the standard bases for irreducible
spinors and tensors of the su(2) algebra used as spin wavefunctions
throughout the paper.  We use spin wavefunctions of the
Rarita-Schwinger type \cite{rar41}, which carry only vector indices
varying from 1 to 3 and spinor indices from 1 to 2, and whose
components are linearly related by symmetry and tracelessness and, for
spinors, by transversality conditions.  In fact, the spinors discussed
in sect.\ \ref{sec:spinor} are the non-relativistic version of
Rarita-Schwinger spinors.  The standard bases are defined so that they
are orthonormal, have definite complex-conjugation properties and
satisfy the Condon--Shortley phase conventions
\cite{wig59,edm96,bie85a,gal90}. 

Our starting point is the usual basis for Pauli spinors, 
\begin{equation}
  \label{eq:spinors}
  \vchi{A}(1/2) = \left(\begin{array}{c} 1\\0\end{array}\right)~,
  \quad
  \vchi{A}(-1/2) = \left(\begin{array}{c} 0\\1\end{array}\right)~.
\end{equation}
The triplet component of basis-spinors tensor products yields a basis
for the representation space of the adjoint representation,
\begin{equation}
  \label{eq:adj}
  \ver{\chi}_{AB}(m) = \sum_{s'_z,s_z=-1/2}^{1/2}
  \CG{\frac{1}{2}}{s'_z}{\frac{1}{2}}{s_z}{1}{m} 
  \ver{\chi}(s_z')_A \epsilon_{BC} \ver{\chi}(s_z)_C~,
  \quad
  m=1,2,3~,
\end{equation}
whose associated unit vectors are,
\begin{equation}
  \label{eq:vects}
  \veps{k}(1) =-\frac{1}{\sqrt{2}} \left(\begin{array}{c}
      1\\i\\0\end{array}\right)~,
  \quad
  \veps{k}(0) =\left(\begin{array}{c} 0\\0\\1\end{array}\right)~,
  \quad
  \veps{k}(-1) =\frac{1}{\sqrt{2}} \left(\begin{array}{c}
      1\\-i\\0\end{array}\right)~.
\end{equation}
These are the standard basis vectors, satisfying,
\begin{equation}
  \label{eq:vects2}
  \veps{k}(s_z)^* = (-1)^{s_z}\veps{k}(-s_z)~,
  \quad
  \veps{k}(s_z)^*\veps{k}(s'_z) = \delta_{s_zs'_z}~,
  \quad
  \sum_{s_z=-1}^1   \veps{k}(s_z)\veps{k'}(s_z)^* = \delta^{kk'}~.
\end{equation}
Furthermore, from the action of the spin operator on basis spinors we
get
\begin{equation}
  \label{eq:1halfpolar}
    \veps{k}(m)\sigma^k_{AB}\vchi{B}(s_z) =
    \frac{2s_z-m}{\sqrt{1+m^2}}\vchi{A}((-1)^m s_z)~,
    \quad
    m=-1,0,1~~,s_z=\pm 1/2~,
\end{equation}
a relation that will be needed in the discussion of spinors
below. 

\subsection{Standard irreducible tensor bases}
\label{sec:tensor}

Starting with the standard vectors (\ref{eq:vects}) we define the
standard basis of irreducible rank-$n$ tensors recursively as,
\begin{equation}
  \label{eq:tensor}
  \veps{i_1\ldots i_n}(m) = \sum_{s_z=-1}^1 \sum_{m'=-n+1}^{n-1}
  \CG{n-1}{m'}{1}{s_z}{n}{m} \veps{i_1\ldots
    i_{n-1}}(m')\veps{i_n}(s_z)~, 
  \quad
  -n\leq m \leq n~.
\end{equation}
This recursion can be solved to yield,
\begin{subequations}
  \label{eq:tensor2}
\begin{equation}
  \label{eq:tensor2a}
  \veps{i_1\ldots i_n}(m) = \sum_{\substack{s_1,\ldots,s_n=-1\\
      s_1+\ldots+s_n=m}}^1
  f_n(s_1,\ldots,s_n) \veps{i_1}(s_1)\ldots\veps{i_n}(s_n)~,
\end{equation}
with
\begin{equation}
  \label{eq:tensor2b}
  f_n(s_1,\ldots,s_n) = \prod_{j=2}^n
  \CG{1}{s_j}{j-1}{\sum_{i=1}^{j-1} s_i}{j}{\sum_{h=1}^{j}s_h}
  = \left(\frac{2^n}{(2n)!}
  \frac{(n+\sum_{i=1}^{n}s_i)!(n-\sum_{j=1}^{n}s_j)!}{\prod_{h=1}^n
    (1+s_h)!(1-s_h)!}\right)^\frac{1}{2}~.
\end{equation}
\end{subequations}
The second equality in (\ref{eq:tensor2b}) follows from the first
one by using the explicit expression for CG coefficients coupling
angular momenta differing in one unit \cite{ham89,gal90}.  From
(\ref{eq:tensor2}) it is clear that $\veps{i_1\ldots i_n}(m)$ is
totally symmetric.  Explicit evaluation shows that $\veps{i_1i_2}(m)$,
$-2\leq m\leq 2$, are traceless and therefore it follows from
(\ref{eq:tensor}) by induction that $\veps{i_1\ldots i_n}(m)$ are
totally traceless for all $n\geq 2$, $-n\leq m\leq n$. Thus,
(\ref{eq:tensor}) defines a set of $2n+1$ irreducible tensors of rank
$n$.  From (\ref{eq:vects2}) and (\ref{eq:tensor}) we obtain by
induction the orthonormality and complex-conjugation relations
\begin{equation}
  \label{eq:tensor3}
  \veps{i_1\ldots i_n}(s_z)^* = (-1)^{s_z}\veps{i_1\ldots i_n}(-s_z)~,
  \quad\quad
  \veps{i_1\ldots i_n}(s_z)^*\veps{i_1\ldots i_n}(s'_z) = \delta_{s_zs'_z}~.
\end{equation}
The completeness relation for the basis of standard tensors defines
the orthogonal projector $X^{i_1\ldots i_n;j_1\ldots j_n}$ in the
space of rank-$n$ tensors onto the subspace of completely symmetric
and traceless tensors
\begin{subequations}
  \label{eq:tensorcomplete}
\begin{equation}
  \label{eq:tensorcompletea}
X^{i_1\ldots i_n;j_1\ldots j_n} = \sum_{s_z=-n}^n \veps{i_1\ldots i_n}(s_z)
\veps{j_1\ldots j_n}(s_z)^*  = \frac{1}{n!} \delta^{\{i_1 j_1} \cdots
\delta^{i_n j_n\}'_0}~.
\end{equation}
Here, the prime on the r.h.s.\ denotes symmetrization over the indices
$i_1\ldots i_n$ and $j_1\ldots j_n$ separately, for example,
\begin{equation}
  \label{eq:tensorcompleteb}
  \begin{gathered}
X^{i;j} = \delta^{ij}~,
\qquad
X^{i_1 i_2;j_1 j_2} = \frac{1}{2} \left(
  \delta^{i_1j_1}\delta^{i_2j_2} + \delta^{i_1j_2}\delta^{i_2j_1} - 
\frac{2}{3} \delta^{i_1i_2}\delta^{j_1j_2} \right)~,\\
\begin{aligned}
  X^{i_1 i_2 i_3;j_1 j_2 j_3} &= \frac{1}{6} \sum_{\sigma\{1,2,3\}}
  \left[ \rule{0pt}{15pt}
\delta^{i_1j_{\sigma_1}}\delta^{i_2j_{\sigma_2}}\delta^{i_3j_{\sigma_3}}
\right. \\
&-\delta^{i_1i_2}\left(
\frac{2}{5} \delta^{j_{\sigma_1}j_{\sigma_2}}\delta^{i_3j_{\sigma_3}}  
-\frac{1}{10} \delta^{j_{\sigma_2}j_{\sigma_3}}\delta^{i_3j_{\sigma_1}}
-\frac{1}{10}\delta^{j_{\sigma_1}j_{\sigma_3}}\delta^{i_3j_{\sigma_2}}
\right) \\
&-\delta^{i_1i_3}\left(
\frac{2}{5} \delta^{j_{\sigma_1}j_{\sigma_3}}\delta^{i_2j_{\sigma_3}}  
-\frac{1}{10} \delta^{j_{\sigma_2}j_{\sigma_3}}\delta^{i_2j_{\sigma_1}}
-\frac{1}{10} \delta^{j_{\sigma_1}j_{\sigma_2}}\delta^{i_2j_{\sigma_3}} 
\right)\\
&-\left. \delta^{i_2i_3}\left(
\frac{2}{5} \delta^{j_{\sigma_2}j_{\sigma_3}}\delta^{i_1j_{\sigma_1}}  
-\frac{1}{10} \delta^{j_{\sigma_1}j_{\sigma_3}}\delta^{i_1j_{\sigma_2}}
-\frac{1}{10} \delta^{j_{\sigma_1}j_{\sigma_2}}\delta^{i_1j_{\sigma_3}} 
\right)\right]~,
\end{aligned}
  \end{gathered}
\end{equation}
\end{subequations}
where in the last equality $\sigma\{1,2,3\}$ is the set of
permutations of $\{1,2,3\}$.  Using the recoupling identity
(\ref{eq:6j-1}) and (\ref{eq:tensor}) we can derive the important
relation,
\begin{equation}
  \label{eq:tensorCG}
    \veps{i_1\ldots i_n}(s_z) = \sum_{s_z=-q}^q \sum_{m'=-n+q}^{n-q}
  \CG{n-q}{m'}{q}{s_z}{n}{m} \veps{i_1\ldots
    i_{n-q}}(m')\veps{i_{n-q+1}\ldots i_n}(s_z)~,
  \quad
  1\leq q \leq n-1~,
\end{equation}
showing that the maximal-rank coupling of two standard tensors is
again a standard basis tensor. From (\ref{eq:tensor2}) we obtain the
equality
\begin{equation}
  \label{eq:tensor2f}
\veps{i_1\ldots i_n}(\pm n) =  \veps{i_1}(\pm1)\cdots \veps{i_n}(\pm1)~,
\end{equation}
which is useful to evaluate reduced matrix elements. We adopt
the convention
\begin{equation}
  \label{eq:tensor0}
  \veps{i_1\ldots i_n}(s_z) = 0 \quad \text{if} \quad |s_z|>n~.
\end{equation}
The matrix $\vec{S}_{(n)}$ of the spin operator $\vec{S}$ in the basis
of spin-$n$ states with $n$ integer is given by
\begin{equation}
  \label{eq:tensorspin}
  \begin{aligned}
  \langle n,s'_z | S^j | n,s_z \rangle &= \veps{i_1\ldots i_n}(s'_z)^*
  (S^j_{(n)})^{i_1\ldots i_n; k_1\ldots k_n} \veps{k_1\dots
    k_n}(s_z)~,\\
(S^j_{(n)})^{i_1\ldots i_n; k_1\ldots k_n} &= \sum_{q=1}^n i
\epsilon^{i_q j k_q} \prod_{\substack{p=1\\p\neq q}}^n \delta^{i_pk_n}~.
  \end{aligned}
\end{equation}

If $A^{i_1\ldots i_n}$ is a rank-$n$ real tensor, or a self-adjoint
tensor operator, its irreducible component is
\begin{equation}
  \label{eq:tensor4}
  \frac{1}{n!} A^{\{i_1\ldots i_n\}_0} = \sum_{s_z=-n}^n \veps{i_1\ldots
    i_n}(s_z) \veps{j_1\ldots j_n}(s_z)^* A^{j_1\ldots j_n} = \sum_{s_z=-n}^n \veps{i_1\ldots
    i_n}(s_z)^* \veps{j_1\ldots j_n}(s_z) A^{j_1\ldots j_n}~.
\end{equation}
Furthermore, for self-adjoint tensor operators the WE
theorem \cite{gal90} holds,
\begin{equation}
  \label{eq:wigeck}
  \langle \ell', \ell'_z | \veps{i_1\ldots i_n}(s_z) A^{i_1\ldots
    i_n} | \ell, \ell_z \rangle =  
  \langle \ell' || \veps{i_1\ldots i_n} A^{i_1\ldots i_n} ||
  \ell\rangle \CG{\ell}{\ell_z}{n}{s_z}{\ell'}{\ell'_z}~.
\end{equation}

For a normal vector $\ver{r}$, let $\mathcal{Y}_{nm}(\ver{r}) =
\veps{i_1\ldots i_n}(m) \ver{r}^{\,i_1}\ldots \ver{r}^{\,i_n}$.  It
follows from (\ref{eq:tensor}) by induction that
\begin{equation}
  \label{eq:tensor5}
  {\vec{L}}^2 \mathcal{Y}_{nm}(\ver{r}) =
  n(n+1) \mathcal{Y}_{nm}(\ver{r})
  \quad \text{and} \quad
  L^3 \mathcal{Y}_{nm}(\ver{r}) = m \mathcal{Y}_{nm}(\ver{r})~.
\end{equation}
We omit the proofs for brevity.  We must then have
$\mathcal{Y}_{nm}(\ver{r}) \propto Y_{nm}(\ver{r})$.  Evaluation of
both sides of the proportionality relation at $\ver{r}=\ver{z}$ yields
\begin{equation}
  \label{eq:tensor6}
   Y_{\ell m}(\ver{r})= 
  \frac{1}{2\sqrt{\pi}} \sqrt{\frac{(2\ell+1)!!}{\ell!}} \veps{i_1\ldots
    i_\ell}(m) \ver{r}^{\,i_1}\ldots \ver{r}^{\,i_\ell}~. 
\end{equation}
From this relation and the addition theorem for spherical harmonics we
get
\begin{equation}
  \label{eq:tensor7}
\begin{aligned}
  \ver{r}^{\,(i_1}\ldots\ver{r}^{\,i_\ell)_0}
  \ver{r}\,'^{\,(i_1}\ldots\ver{r}\,'^{\,i_\ell)_0} &= \sum_{m=-\ell}^\ell
  \veps{i_1\ldots i_\ell}(m) \veps{j_1\ldots j_\ell}(m)^* 
  \ver{r}^{\,i_1}\ldots\ver{r}^{\,i_\ell}
  \ver{r}\,'^{\,j_1}\ldots\ver{r}\,'^{\,j_\ell}\\
  &=
  \frac{\ell!}{(2\ell-1)!!} P_\ell(\ver{r}\cdot\ver{r}\,')~.
\end{aligned}
\end{equation}
The relation (\ref{eq:tensor6}) can be inverted, for each $\ell$, by
extending the basis (\ref{eq:tensor}) with appropriate reducible
tensors to an orthogonal basis of the entire space of rank-$n$
tensors.

\subsection{Standard spinor bases}
\label{sec:spinor}

We define the standard spin-$(n+1/2)$ spinors with $n\geq 1$ as
\begin{equation}
  \label{eq:nspinor}
  \vchiU{A}{i_1\ldots i_n}(s_z) = \sum_{s'_z=-1/2}^{1/2}\sum_{m=-n}^n
  \CG{n}{m}{1/2}{s'_z}{n+1/2}{s_z} \veps{i_1\ldots i_n}(m) \vchi{A}(s'_z)~,
  \qquad
  -n-1/2 \leq s_z \leq n+1/2~.
\end{equation}
Using (\ref{eq:tensor}) and the recoupling identity (\ref{eq:6j-1}), 
from (\ref{eq:nspinor}) we find the equivalent expressions,
\begin{equation}
    \label{eq:nspinor1}
  \vchiU{A}{i_1\ldots i_n}(s_z) =
  \sum_{s'_z=-q-1/2}^{q+1/2}\sum_{m=-n+q}^{n-q}    
  \CG{n-q}{m}{q+1/2}{s'_z}{n+1/2}{s_z} \veps{i_1\ldots i_{n-q}}(m)
  \vchiU{A}{i_{n-q+1}\ldots i_n}(s'_z)~,
\end{equation}
with $1\leq q \leq n-1$.
From definition (\ref{eq:nspinor}) we see that $\vchiU{A}{i_1\ldots
  i_n}$ is completely symmetric and traceless in its tensor indices.
Thus, (\ref{eq:nspinor1}) remains valid after
an arbitrary permutation of tensor indices in its r.h.s.  From
(\ref{eq:nspinor}) with $n=1$ and (\ref{eq:1halfpolar}) we find, for
the spin-3/2 spinor, $\sigma^i_{AB} \vchiU{B}{i}(s_z)=0$, $-3/2
\leq s_z \leq 3/2$.  Thus, from (\ref{eq:nspinor1}) with $q=1$ we
immediately obtain,
\begin{equation}
  \label{eq:nhalfpolar}
\sigma^{i_k}_{AB} \vchiU{B}{i_1\ldots i_k\ldots i_n}(s_z)=0~,
\qquad
1\leq k\leq n~,
\quad
-n-1/2 \leq s_z \leq n+1/2  ~.
\end{equation}
From (\ref{eq:nspinor}) and (\ref{eq:nhalfpolar}), the number of
independent components in the spinor basis is $2n+2$ as it should.
The orthornormality and complex-conjugation relations read
\begin{equation}
  \label{eq:ortoconj}
  \vchiU{A}{i_1\ldots i_n}(s_z)^*  \vchiU{A}{i_1\ldots i_n}(s'_z)=\delta_{s_zs'_z}~,
  \qquad
  i\sigma^2_{AB} \vchiU{B}{i_1\ldots i_n}(s_z) =
  (-1)^{1/2+s_z} \vchiU{A}{i_1\ldots i_n}(-s_z)^*~.
\end{equation}
The completeness relation for spin-$s$ spinors, with $s=n+1/2$, $n\geq
1$, defines the orthogonal projector $X^{i_1\ldots i_n;j_1\ldots
  j_n}_{AB}$ onto the subspace of spin-$(n+1/2)$ spinors of
$\mathbb{C}^{(2n+1)\times 2}$, 
\footnote{For notational simplicity we identify the space of complex
  rank-$n$ completely  symmetric and traceless tensors with
  $\mathbb{C}^{(2n+1)}$, which has the same dimension.} 
\begin{equation}
  \label{eq:nspincompl}
  \begin{aligned}
  \lefteqn{X^{i_1\ldots i_n;j_1\ldots j_n}_{AB} \equiv \sum_{s_z=-s}^{s}
  \vchiU{A}{i_1\ldots i_n}(s_z)
  \vchiU{B}{j_1\ldots j_n}(s_z)^*} \\
  &=X^{i_1\ldots i_n;k_1\ldots k_n} \left[
    \frac{n+1}{2n+1} \delta^{k_1h_1}\ldots \delta^{k_nh_n}
    \delta_{AB}+
    \frac{n}{2n+1} i \epsilon^{k_1rh_1} \sigma^{r}_{AB}
    \delta^{k_2h_2}\ldots \delta^{k_nh_n}  \right]
  X^{i_1\ldots i_n;k_1\ldots k_n}~,
  \end{aligned}
\end{equation}
where $X^{i_1\ldots i_n;j_1\ldots j_n}$ is the projector defined in
(\ref{eq:tensorcomplete}).  The orthogonal complement of the
spin-$(n+1/2)$ subspace in 
$\mathbb{C}^{(2n+1)\times 2}$ is $\ker(X)$,
which can be parameterized as 
$\sigma^{\{i_1}_{AB} \vchiU{B}{i_2\ldots i_n\}_0}$ 
with
$\vchiU{A}{i_1\ldots i_{(n-1)}}$
a spin-$(n-1/2)$ spinor.  This is the spin-$(n-1/2)$ subspace of
$\mathbb{C}^{(2n+1)\times 2}$. 

The spin operator for $s=(n+1/2)$ spinors is given by the operator
tensor product of the spin-$n$ and spin-1/2 operators, its matrix
being 
\begin{equation}
  \label{eq:spinop}
  ( S^j_{(n+1/2)} )^{i_1\ldots i_n;k_1\ldots k_n}_{AB} =
  ( S^j_{(n)} )^{i_1\ldots i_n;k_1\ldots k_n}\delta_{AB} +
  \frac{1}{2}\delta^{i_1k_1}\ldots \delta^{i_nk_n}\sigma^j_{AB}~, 
\end{equation}
with $S^j_{(n)}$ defined in (\ref{eq:tensorspin}).
From (\ref{eq:nhalfpolar}) and (\ref{eq:spinop}) we obtain the
following useful identities, 
\begin{gather}
  \label{eq:nhalfid}
  i\epsilon^{i_1kh}\sigma^k_{AB} \vchiU{B}{hi_2\ldots i_n}(s_z) =
  \vchiU{A}{i_1\ldots i_n}(s_z)~,\\
  \sigma^k_{AB} \vchiU{B}{i_1\ldots i_n}(s_z) - \sigma^{i_1}_{AB}
  \vchiU{B}{ki_2\ldots i_n}(s_z) + i
  \epsilon^{ki_1h} \vchiU{A}{hi_2\ldots i_n}(s_z) = 0~,\\
  \vchiU{A}{i_1\ldots i_{n-1}h}(s'_z)^* \sigma^k_{AB}
  \vchiU{B}{j_1\ldots j_{n-1}h}(s_z) =
  \vchiU{A}{i_1\ldots i_{n-1}p}(s'_z)^* i\epsilon^{pkq}
  \vchiU{A}{j_1\ldots j_{n-1}q}(s_z)~,\\ 
  \begin{aligned}
  \vchiU{A}{i_1\ldots i_n}(s'_z)^* (S^k_{(n+1/2)})^{i_1\ldots
    i_n;j_1\ldots j_n}_{AB} \vchiU{B}{j_1\ldots j_n}(s_z) &=
  (n+1/2) \vchiU{A}{h_1\ldots h_n}(s'_z)^* \sigma^k_{AB}
  \vchiU{B}{h_1\ldots h_n}(s_z) \\
  &=
  (n+1/2) \vchiU{A}{h_1\ldots h_{n-1}p}(s'_z)^* i\epsilon^{pkq} 
  \vchiU{B}{h_1\ldots h_{n-1}q}(s_z) ~,
  \end{aligned}
\end{gather}
which remain valid after permutation of $i_1\ldots i_n$, etc.

\setcounter{equation}{0}
\section{Reduced matrix elements}
\label{sec:matele}

In this appendix we gather expressions for matrix elements of certain
irreducible tensor operators needed in the foregoing.  The notation used
is the same as in (\ref{eq:wigeck}).  We begin from the known results
\cite{gal90} 
\begin{align}
  \label{eq:redmat1}
  \langle j'||\veps{i} J^i ||j\rangle &= \sqrt{j(j+1)} \delta_{jj'}~,\\
  \langle \ell' || Y_{L} ||\ell\rangle &= \frac{1}{\sqrt{4\pi}}
    \sqrt{\frac{(2\ell+1)(2L+1)}{2\ell'+1}}
  \CG{\ell}{0}{L}{0}{\ell'}{0}~.   \label{eq:redmat2}
\end{align}
In (\ref{eq:redmat1}) $J^i$ can of course be any angular momentum
operator: orbital $L^i$, spin $S^i$ or total $J^i$.  For tensor
products of angular momentum operators we have, using
(\ref{eq:tensor2f}) and (\ref{eq:redmat1}), 
\begin{align}
  \langle j'||\veps{i_1\ldots i_n} J^{i_1}\cdots J^{i_n} ||j\rangle =
  \frac{n!}{\sqrt{2^n (2n)!}}\frac{1}{\sqrt{2j+1}} \sqrt{\frac{(2j+n+1)!}{(2j-n)!}}
  \,\delta_{jj'}~.  \label{eq:redmat3}
\end{align}
For convenience, we quote here the explicit form of the CG coefficient
appearing in (\ref{eq:redmat2}).  If $t$ is odd the CG coefficient
vanishes, and if $t\ge 0$ is even
\begin{equation}
  \label{eq:redmat4}
  \begin{split}
  \CG{\ell}{0}{n+t}{0}{\ell_n}{0} &= (-1)^{t/2} (\ell_1-\ell)^n 
\sqrt{\frac{(2n+t-1)!! (t-1)!!}{(n+t/2)!(t/2)!}} \sqrt{2\ell_n+1} 
\sqrt{\frac{((\ell_n+\ell+n+t)/2)!}{((\ell_n+\ell-n-t)/2)!}}\\
&\qquad\times
\sqrt{\frac{(\ell_n+\ell-n-t-1)!!}{(\ell_n+\ell+n+t+1)!!}}~,
\qquad \text{$t\ge 0$ even}.
  \end{split}
\end{equation}
Next, we consider tensor products of the position versor
$\ver{r}=\vec{r}/r$.  From (\ref{eq:tensor6}) and (\ref{eq:redmat2})
\begin{equation}
  \label{eq:redmat5}
  \langle \ell'||\veps{i_1\ldots i_n} \verr{i_1}\ldots\verr{i_n}
  ||\ell\rangle =
  \sqrt{\frac{n!}{(2n-1)!!}}\sqrt{\frac{2\ell+1}{2\ell'+1}}
  \CG{\ell}{0}{n}{0}{\ell'}{0}  ~.
\end{equation}
In the case $|\Delta\ell|=n$, which is of interest to us, a more
explicit expression is given by,
\begin{equation}
  \label{eq:redmat6}
  \begin{aligned}
\langle\ell_n||\veps{i_1\ldots
  i_n}\verr{i_1}\ldots\verr{i_n}||\ell\rangle &= 
\frac{(-1)^\frac{n-\Delta\ell}{2}}{2^{n/2}}
\sqrt{\frac{2\ell+1}{2(\lave+\frac{n}{2})+1}}  
\sqrt{\frac{\Gamma\left(\lave - \frac{n}{2}+\frac{1}{2}\right)}
{\Gamma\left(\lave + \frac{n}{2}+\frac{1}{2}\right)}}\\
&\quad \times \sqrt{\frac{\Gamma\left(\lave +
      \frac{n}{2}+1\right)} {\Gamma\left(\lave - 
    \frac{n}{2}+1\right)}}~.
  \end{aligned}
\end{equation}
The cases $n=1,$ 2, 3 are of particular importance to us.  Noticing
that $(-1)^{\frac{|\dl|-\dl}{2}} = \left(
  \frac{\dl}{|\dl|}\right)^{|\dl|}$, we get
\begin{subequations}
  \label{eq:redmat6aux}
  \begin{align}
  \label{eq:redmat6auxa}
    \langle\ell_1||\veps{i}\verr{i}||\ell\rangle &=
    \frac{\dl}{\sqrt{2}} \sqrt{\frac{2\lave+1}{2\ell_1+1}}~,\\
  \label{eq:redmat6auxb}
   \langle\ell_2||\veps{i_1i_2}\verr{i_1}\verr{i_2}||\ell\rangle &= 
   \sqrt{2\ell+1} \sqrt{\frac{\lave (\lave+1)}{(2\lave-1)(2\lave+1)
     (2\lave+3)}}~,\\
  \label{eq:redmat6auxc}
   \langle\ell_3|| \veps{i_1i_2i_3} \verr{i_1} \verr{i_2} \verr{i_3}
   ||\ell\rangle &= \frac{\dl}{24\sqrt{2}}
   \sqrt{\frac{2\ell+1}{\lave+2}}
   \sqrt{\frac{(2\lave-1)(2\lave+1)(2\lave+3)}
     {(\lave-1)\lave(\lave+1)}}~.
  \end{align}
\end{subequations}
Equations (\ref{eq:redmat6}) and (\ref{eq:redmat6aux}) remain valid
under the replacements $\ell\rightarrow s$, $\verr{i}\rightarrow T^i$,
that give the reduced matrix elements $\langle s\pm n||\veps{i_1\ldots
  i_n}T^{i_1}\ldots T^{i_n}||s\rangle$ of tensor products of the spin
transition operator for integer or half-integer $s$.

For tensor products of $\ver{r}$ and $\vec{L}$, using
(\ref{eq:redmat3}), (\ref{eq:redmat6}) and (\ref{eq:tensor2f}) we
obtain, 
\begin{equation}
  \label{eq:redmat7}
  \begin{aligned}
\lefteqn{
\langle\ell+\Delta\ell || \veps{i_1\ldots
  i_n}\verr{i_1}\ldots\verr{i_{|\Delta\ell|}}
L^{i_{|\Delta\ell|+1}}\ldots L^{i_n}||\ell\rangle = 
\left(\frac{\dl}{|\dl|}\right)^{|\dl|} 
\frac{1}{2^{n/2}2^{|\Delta\ell|}} 
\sqrt{\frac{(n+|\Delta\ell|)!(n-|\Delta\ell|)!}{(2n)!}}
}\hspace{35ex}
 \\
&\times \frac{\sqrt{2\ell+1}}{2\lave+|\Delta\ell|+1}
\sqrt{\frac{(2\lave+n+1)!}{(2\lave-n)!}} 
\frac{\Gamma\left(\lave - |\Delta\ell|/2 +
    1/2\right)}
{\Gamma\left(\lave + |\Delta\ell|/2 + 1/2\right)}~. 
  \end{aligned}
\end{equation}
This equation reduces to (\ref{eq:redmat3}) for $\Delta\ell=0$, and to
(\ref{eq:redmat6}) for $|\Delta\ell|=n$.    The particular
cases $n=2$, 3 of (\ref{eq:redmat7})  are important in the
foregoing,
\begin{subequations}
\label{eq:redmat8}
\begin{align}
  \langle \ell_1||\veps{ij} \verr{i}L^j ||\ell\rangle &=
  \frac{\Delta\ell}{4}\frac{1}{\sqrt{2\ell_1+1}}
  \sqrt{(2\lave-1)(2\lave+1)(2\lave+3)}~,
  \label{eq:redmat8a}   \\
  \langle \ell_1||\veps{ijk} \verr{i}L^j L^k ||\ell\rangle &=
  \frac{\Delta\ell}{2\sqrt{30}}\frac{1}{\sqrt{2\ell_1+1}}
  \sqrt{(2\lave-2)(2\lave-1)(2\lave+1)(2\lave+3)(2\lave+4)}~,
  \label{eq:redmat8b}   \\
  \langle \ell_2||\veps{ijk} \verr{i}\verr{j} L^k ||\ell\rangle &=
  \frac{1}{\sqrt{3}} \sqrt{2\ell+1} \sqrt{\frac{(\lave-1) \lave
      (\lave+1) (\lave+2)}{(2\lave-1) (2\lave+1) (2\lave+3)}}~.
  \label{eq:redmat8c}
\end{align}
\end{subequations}
The following mixed reduced matrix elements are also useful when
dealing with tensor decompositions in irreducible components,
\begin{subequations}
\label{eq:redmat9}  
\begin{align}
  \langle \ell_1||\veps{i} (\ver{r}\wedge \vec{L})^i ||\ell\rangle &=
  -\frac{i}{2\sqrt{2}}\left((\ell_1+\ell+1)-2\dl\right)
    \sqrt{\frac{\ell_1+\ell+1}{2\ell_1+1}}~, \label{eq:redmat9a}\\
  \langle \ell_1||\veps{ij} (\ver{r}\wedge \vec{L})^i L^j||\ell\rangle &= 
  -\frac{i}{8}\left((\ell_1+\ell+1)-2\dl\right)
   \sqrt{\frac{(\ell_1+\ell-1)(\ell_1+\ell+1)(\ell_1+\ell+3)}{2\ell_1+1}}
  ~, \label{eq:redmat9b}\\
  \langle \ell_2||\veps{ij} \verr{i} (\ver{r}\wedge\vec{L})^j ||\ell\rangle &=
  -\frac{i}{2}
  \left((\ell_1-\ell)(2\ell_1+1)-3\right)
\sqrt{\frac{\ell_1(\ell_1+1)}{(2\ell_1+1)(2\ell_2+1)}}~,
  \label{eq:redmat9c}
\end{align}
\end{subequations}
With the replacements $\ell\rightarrow s$, $\Delta\ell\rightarrow
\Delta s$, $\verr{i}\rightarrow T^i$, $L^i\rightarrow S^i$,
(\ref{eq:redmat7}) and (\ref{eq:redmat8}) give the reduced matrix
elements $\langle s\pm n||\veps{i_1\ldots i_n}T^{i_1}\ldots
T^{i_{|\Delta s|}} S^{i_{|\Delta s|+1}}\ldots S^{i_n}||s\rangle$ for
integer or half-integer $s$, and similarly for (\ref{eq:redmat9}).  

\setcounter{equation}{0}
\section{The matrix elements $\boldsymbol{\langle
\verpp{}|\P_{\ell_n}\verr{i_1}\ldots\verr{i_n}\P_\ell 
| \verp{}\rangle}$}
\label{sec:appa}

In this appendix we derive (\ref{eq:rmat1}) and
(\ref{eq:rmat2}). First, we consider the matrix element,
\begin{equation}
  \label{eq:appa1}
  \begin{split}
  \langle
\verpp{}|\P_{\ell_1}\verpp{}\cdot\verr{}\P_\ell |\verp{}\rangle &=
\int d^2q\,   \langle \verpp{}|\P_{\ell_1}|\verq{}\rangle
\verpp{}\cdot\verq{} 
\langle \verq{}|\P_{\ell}|\verp{}\rangle\\
&= \frac{2\ell_1+1}{4\pi} \frac{2\ell+1}{4\pi}
\int d^2q\,   P_{\ell_1}(\ppdotq) \ppdotq P_{\ell}(\pdotq)\\
&= \frac{\ell_1+1}{2\ell_1+3} \langle \verpp{}|\P_{\ell_1+1}
\P_{\ell}|\verp{}\rangle +
\frac{\ell_1}{2\ell_1-1} \langle \verpp{}|\P_{\ell_1-1}
\P_{\ell}|\verp{}\rangle~,
  \end{split}
\end{equation}
where the integrals extend over the unit sphere, and in the last
equality we used the relation
\begin{equation*}
  P_{\ell_1}(x) = \frac{\ell_1+1}{2\ell_1+1} P_{\ell_1+1}(x) +
  \frac{\ell_1}{2\ell_1-1} P_{\ell_1-1}(x)~.
\end{equation*}
Applying the projector property of $\P_\ell$ in the last line of
(\ref{eq:appa1}) we obtain,
\begin{equation}
  \label{eq:appa2}
\langle \verpp{}|\P_{\ell_1}\verpp{}\cdot\verr{}\P_\ell
|\verp{}\rangle =
\frac{1}{2}\frac{1}{2\ell+1}(\ell_1+\ell+1)
\langle
\verpp{}| \P_\ell |\verp{}\rangle~.
\end{equation}
Similarly, using 
\begin{equation*}
  \nabla_{p'} \ppdotq = \verpp{}\wedge(\verq{}\wedge\verpp{})~,
\end{equation*}
we get,
\begin{equation}
  \label{eq:appa3}
\langle \verpp{}|\P_{\ell_1}\verpp{}\wedge(\verr{}\wedge \verpp{})\P_\ell
|\verp{}\rangle =  
\nabla_{p'} \left(\frac{\ell_1-\ell}{4\pi} P_{\ell_1}(\pdotpp)\right)
= -\frac{\ell_1-\ell}{4\pi} \verpp{}\wedge(\verp{}\wedge \verpp{})
P'_\ell(\pdotpp)~. 
\end{equation}
Summing (\ref{eq:appa2}) and (\ref{eq:appa3}) we obtain
(\ref{eq:rmat2}) with the help of the relation
\begin{equation}
  \label{eq:appa4}
  x P'_{\ell_k}(x) - (n-k) P_{\ell_k}(x) + \frac{1}{2}
  (\ell_{n+1}-\ell_n) (\ell_{n+1} + \ell_n +1) P_{\ell_k} =
  P'_{\ell_{k+1}}~, 
  \quad
  0\leq k \leq n~.
\end{equation}
The derivation of (\ref{eq:rmat1}) runs along the same lines,
\begin{equation}
  \label{eq:appa5}
  \begin{aligned}
\lefteqn{
\langle \verpp{}|\P_{\ell_{n+1}}\verr{i_1}\ldots \verr{i_n}\verr{i_{n+1}}\P_\ell
|\verp{}\rangle =    
\int d^2q\, 
\langle \verpp{}|\P_{\ell_{n+1}}\verr{i_{n+1}}\P_{\ell_n} |\verq{}\rangle 
\langle \verq{}|\P_{\ell_{n}}\verr{i_1}\ldots \verr{i_n}\P_\ell
|\verp{}\rangle} \hspace{25ex}\\
&= \int d^2q\, \langle
\verpp{}|\P_{\ell_{n+1}}\verpp{i_{n+1}}\verpp{}\cdot
\verr{}\P_{\ell_n} |\verq{}\rangle \langle
\verq{}|\P_{\ell_{n}}\verr{i_1}\ldots \verr{i_n}\P_\ell
|\verp{}\rangle\\
&+ \int d^2q\, \langle \verpp{}|\P_{\ell_{n+1}}(\verr{i_{n+1}} -
\verpp{i_{n+1}}\verpp{}\cdot \verr{})\P_{\ell_n} |\verq{}\rangle
\langle \verq{}|\P_{\ell_{n}}\verr{i_1}\ldots \verr{i_n}\P_\ell
|\verp{}\rangle~.
  \end{aligned}
\end{equation}
Applying now (\ref{eq:appa2}) to the first integral and
(\ref{eq:appa3}) to the second one in the last equality, we obtain
(\ref{eq:rmat1}).


\begin{thebibliography}{99}
%
\bibitem{edm96} A.\ R.\ Edmonds, ``Angular Momentum in Quantum
  Mechanics,'' Princeton Univ.\ Press, Princeton, 1996.
%
\bibitem{wig59} E.\ P.\ Wigner, ``Group Theory and its Application to
  the Quantum Mechanics of Atomic Spectra,'' Academic Press, New York,
  1959.
%
\bibitem{jud75} B.\ R.\ Judd, ``Angular Momentum Theory for Diatomic
  Molecules,'' Academic Press, New York, 1975.
%
\bibitem{bie85a} L.\ C.\ Biendenharn, J.\ D.\ Louck, ``Angular Momentum
  in Quantum Physics,'' Encyclopedia of Mathemathics and its
  Applications Vol.\ 8, Cambridge Univ.\ Press, New York, 1985.
%
\bibitem{bie85b} L.\ C.\ Biendenharn, J.\ D.\ Louck, ``The Racah-Wigner
  algebra in Quantum Theory,'' Encyclopedia of Mathemathics and its
  Applications Vol.\ 9, Cambridge Univ.\ Press, New York, 1985.
%
\bibitem{var88} D.\ A.\ Varshalovich, A.\ N.\ Moskalev, V.\ K.\
  Khersonskii, ``Quantum Theory of Angular Momentum,'' World
  Scientific, Singapore, 1988.
%
\bibitem{ham89} M.\ Hamermesh, ``Group Theory and its Application to
  Physical Problems,'' Dover, New York, 1989.
%
\bibitem{gal90} A.\ Galindo, P.\ Pascual, ``Quantum Mechanics,'' Vol.\
  I, Springer--Verlag, New York, 1990.
%
\bibitem{jackxx} J.\ D.\ Jackson, ``Classical Electrodynamics,'' John
Wiley, New York, 1999.
%
\bibitem{bou10} A.\ Bouzas, ``Addition theorems for spin spherical
  harmonics. II Results,'' to be published.
%
\bibitem{nac90} O.\ Nachtmann, ``Elementary Particle Physics,''
  Springer--Verlag, New York, 1990.
%
\bibitem{pdg} Particle Data Group, C.\ Amsler et al., ``Review of
  Particle Physics,'' Phys.\ Lett.\ B\textbf{667}, 2008, 1. 
%
\bibitem{wolf} S.\ Wolfram, ``The Mathematica Book,''  Third Edition,
  Cambridge U.\ Press, New York, 1996.
%
\bibitem{rar41} W.\ Rarita, J.\  Schwinger, Phys.\ Rev.\ \textbf{60},
  1941, 61.
%
\end{thebibliography}
\end{document}